# Coupling between sub-mesoscale eddies, internal waves, and turbulence in the deep Mediterranean: A spectral investigation

by Hans van Haren

Royal Netherlands Institute for Sea Research (NIOZ), P.O. Box 59, 1790 AB Den Burg, the Netherlands.
e-mail: hans.van.haren@nioz.nl

**Abstract.** Interaction between energy-abundant (sub-)mesoscale eddies and internal waves can lead to turbulence generation and may prove important for replenishment of nutrients for deep-sea life and circulation. However, observational evidence of such interaction is scarce and precise energy transfer is unknown. In this paper, an extensive spectral study is reported using mooring data from nearly 3000 high-resolution temperature sensors in about half-a-cubic hectometer of seawater above a deep flat Northwestern-Mediterranean seafloor. The number of independent data records partially improves statistics for better determination of spectral slopes, which however do not show a roll-off to the viscous dissipation range of turbulence. The spectra hardly show power-laws $\omega^p$ having exponent $p = -5/3$ representing an inertial subrange that evidences shear-induced isotropic turbulence. Instead, they are dominated by $p = -7/5$ representing a buoyancy subrange, which evidences convection-induced anisotropic turbulence. In contrast with $p=-5/3$ that indicates a downgradient cascade of energy, $p=-7/5$ characterizes by an ambiguous cascade direction. At height $h<50$ m above seafloor, $p=-7/5$ is found adjacent to instrumental noise. The $p=-7/5$ is also found in the sub-mesoscale/internal wave band that is elevated in variance by one order of magnitude. It is reasoned that this sub-inertial range cannot represent isotropic motions, hence $p \neq -5/3$ at all heights, and a new deep-sea energy cascade is proposed between mesoscales and turbulence dissipation. Only higher up in more stratified waters an inertial subrange is formed. The transition from internal waves into large-scale turbulence follows $p = -2$, while a higher-frequency transition from 0 to $\pi$ phase change reflects overturns of slanted convection or standing-wave breaking leading to isotropic turbulence.

# 1 Introduction

In a stably stratified environment like the sun-heated ocean, downward pulses of warm water seem impossible in terms of irreversible turbulent convection. 'Natural, buoyancy-driven convection' (e.g. Dalziel et al., 2008) applies to denser waters moving down and less dense waters moving up. In terms of temperature variations, this implies cooler waters moving down and warmer up. (An execption can occur when the other major contributor to ocean-density variations dominates over temperature variations: If downward moving warm waters are saltier than their environment, and vice versa for upward motions). In contrast with day-time atmosphere dynamics, in the ocean such turbulent convection may occur regularly under limited conditions: in the upper 10 m near the surface during nighttime (Brainerd and Gregg, 1995), and possibly in the lower 100 m above the seafloor due to general geothermal heating (e.g., Pasquale et al. 1996), depending on the local stratification. It can also occur as 'deep dense water formation' after specific preconditioning near the surface in localized areas like polar seas and the Mediterranean during brief irregular periods of about a week. While occurring during most winters down to several 100's of meters, only once every decade or so newly formed dense waters go down all the way to >2000 m deep seafloors (Rhein, 1995; Dickson, 1996; Mertens and Schott, 1998). Its convection turbulence has thus far not been directly observed in the deep sea (Thorpe, 2005).

One of the potential results, in fact evidence, of convection is the development of two-dimensional '2D' eddies, which also associate with front(al collapse), and which have horizontal scales O(0.1-10) km in the ocean: sub-mesoscales (e.g., Taylor and Thompson, 2023). (While eddies are characterized here as 2D rather than 3D because of their small aspect ratio << 1, it is acknowledged that they generally have stronger flows in more stratified waters near the surface than in the deep sea). Sub-mesoscale eddies are found ubiquitous in the Western Mediterranean (Gascard, 1978; Testor and Gascard, 2006) and are suggested to interact with larger scale features such as continental boundary flows. In this region, other mechanisms may lead to sub-mesoscale eddies as well, when deep dense water is not formed. For example, the same boundary flows may spin off eddies, at mesoscale with typical 10-20 day periodicity (Crepon et al., 1982; van Haren et al., 2011), which may break up to sub-mesoscales.

The region also shows extensive generation of near-inertial waves, which under conditions of weakly stratified waters lead to slantwise convection, as has been proposed in general (Marshall and Schott, 1999; Straneo et al., 2002) and inferred from shipborne profiling observations in the open Western Mediterranean (van Haren and Millot, 2009). The convection may develop to irreversible 3D turbulence, while vertical (opposite-to-gravity) density profiles appear stably stratified. Under such conditions, the traditional internal-wave bounds [f, N] of inertial f = 2Ωsinφ and buoyancy N frequencies extend to inertio-gravity wave IGW bounds [$\omega_{min}$ $\omega_{max}$] (LeBlond and Mysak, 1978). Here, φ denotes latitude, and Ω the Earth rotational frequency. The $\omega_{min}$<f and $\omega_{max}$>2Ω,N, whichever is largest, are functions of N, φ and direction of wave propagation (LeBlond and Mysak, 1978; Gerkema et al., 2008),

$$\omega_{max}, \omega_{min} = (A \pm (A^2-B^2)^{1/2})^{1/2}/\sqrt{2}, \qquad (1)$$

in which A = $N^2 + f^2 + f_s^2$, B = 2fN, and $f_s$ = $f_h$sinα, $f_h$ = 2Ωcosφ and α the angle to φ. For $f_s$ = 0 or N >> 2Ω, the [f, N] are retrieved from (1).

Internal-wave bounds may also vary due to (sub-)mesoscale motions. Local time- and space-varying horizontal waterflow (U, V) differences such as in meanders and eddies can generate relative vorticity ζ = ∂V/∂x - ∂U/∂y, so that the effective inertial frequency reads (Kunze, 1985),

$$f_{eff} = (f^2 + f\zeta - \partial U/\partial x \cdot \partial V/\partial y)^{1/2}. \qquad (2)$$

For weak ζ < 0.2f, (2) can be approximated by $f_{eff} \approx f + \zeta/2$ (Mooers, 1975; Kunze, 1985), but this is probably not relevant for the Northwestern Mediterranean where |ζ| = f/2 are reported from mid-depth drifter observations (Testor and Gascard, 2006). When $f_{eff}$ < f, relative vorticity is dominated by anticyclonic motions, while for $f_{eff}$ > f cyclonic motions dominate.

A reversible, also 3D, process occurs when internal-wave motions displace the stably-stratified environment (e.g., LeBlond and Mysak, 1978). Such motions may displace relatively warm waters downward during a particular wave-phase, and cooler waters up. However, such displacements will not overturn irreversibly and thus not vertically mix different water masses.

A combination of the two 3D processes was observed in fresh-water alpine Lake Garda, where in the weakly stratified waters underneath internal waves turbulent convection appeared (van Haren and Dijkstra, 2021). It was suggested that this convection was either generated via shear displacing convection tubes slantwise, or via wave-accelerations overcoming the density differences in reduced gravity instead of overcoming gravity as in natural convection. However, precise coupling interactions between above 3D motions and turbulence are still unknown, despite historic suggestions (e.g., Ozmidov, 1965a).

A universal surface-ocean [kinetic] energy spectrum was proposed by Ozmidov (1965a). As shown in Fig. 1a, a limited number of peaks representing energy sources like large basin-scale circulation, inertial motions and tides, and wind waves were connected with power-laws of different levels but with a single exponent p = -5/3 representing unidirectional downscale energy cascade (Kolmogorov, 1941).

In this paper, observational investigations of convective water pulses are further pursued that occur frequently in the deep Western Mediterranean. The attempt is to better understand the deep-sea energy cascade including from sources like (sub-)mesoscale motions that were not considered in Ozmidov (1965a). The observations are made with the aid of a large 3D mooring-array holding nearly 3000 high-resolution temperature 'T'-sensors on 45 closely spaced vertical lines (Fig. 1b). Three current meters were added to verify (2) on small horizontal scales of about 50 m. The original aim was to try to capture deep dense-water events, but these were not found in 20-month long records that included two winters.

Here, the focus will be on frequency ($\omega$) spectral analysis, which was motivated by a re-analysis of historic upper-sea current-meter mooring data from the region (van Haren, 2025). While the historic summer data were characterized by a larger near-inertial peak in kinetic energy, the more energetic winter data showed a power-law $\omega^p$ with exponent of about p = -2.2, across sub-mesoscale and internal wave bands. The same slope was found in the summer-data, at sub-inertial frequencies mainly. Coarsely, such a slope shows (unnamed) in the same frequency range from other ocean areas (e.g., van Aken et al., 2005; Ferrari and Wunsch, 2009). As these data did not resolve the turbulence range and generally have poor spectral statistics, renewed investigation is performed using data from the large 3D mooring holding many sensors for expected improved statistics.

## 2 Spectral scaling

Depending on the predictability of the motions, an ocean-spectral peak may be <<1 order of frequency-range wide, e.g. for deterministic barotropic tides, or about one-quarter order wide, e.g. for (baroclinic-internal) inertial waves. Less predictable broader-range distributions of variance may be considered over one or more orders wide, which are modeled by power-laws $\omega^p$ with exponents like p = -5/3.

Since Kolmogorov (1941), a generally accepted scaling model exists for part of the isotropic turbulence spectrum in an inertial subrange of continual irreversible cascade of energy from large to small scales. However, such a consistent model does not exist for energy cascade from (sub-)mesoscales, perhaps via IGW, to turbulence in stratified waters. For the transition from meso- to sub-mesoscales, spectral slopes with exponents varying from $p = -5/3$ to $p = -3$ have been suggested (Ozmidov, 1965a (without source-naming); McWilliams, 2016; Storer et al., 2022). No mention has been made of scaling spectra with a slope proposed by Bolgiano (1959) and Obukhov (1959) 'BO' for active-scalar buoyancy-subrange convection turbulence.

This BO-scaling has different spectral slopes for scalars (potential energy) and kinetic energy, being $p = -7/5$ and $p = -11/5$, respectively. BO-scaling has an ambiguous direction for energy cascade, at least for homogeneous turbulence (Lohse and Xia, 2010). Thus, it should be contrasted with above 'KO'-scaling proposed by Kolmogorov (1941) and refined by Obukhov (1949), for passive-scalar shear-turbulence scales (Tennekes and Lumley, 1972; Warhaft, 2000), so that $p = -5/3$ is identical for both scalars and kinetic energy. Buoyancy-driven BO-scaling associates with anisotropic turbulence, with Richardson number Ri of order unity, and is not existent under neutral [no convection] conditions. It contrasts with KO-scaling, which may become non-existent under sufficiently stable conditions. These notions by Bolgiano (1959) suggest that KO- and BO-scaling cannot co-exist at the same location. The appearance of mean dominant KO- and BO-scaling is investigated for the deep Mediterranean.

Other important exponents of slopes in ocean spectra are $p = -1$ for intermittency 'Im'-scaling of a weakly chaotic nonlinear system (Schuster, 1984; Bak et al., 1987), and $p = -2$ 'IW-scaling' for internal wave (Garrett and Munk, 1972) or finestructure contamination (Phillips, 1971; Reid, 1971).

We elaborate on turbulence likely induced by internal-wave breaking, via parametric instability (Davis and Acrivos, 1967), slantwise convection (Straneo et al., 2002), and/or possibly via coupling with sub-mesoscale eddies (e.g., Chunchuzov et al., 2021).

## 3 Materials and Methods

A nearly half-cubic-hectometer of seawater was sampled using 2925 self-contained high-resolution NIOZ4 temperature (T-)sensors. Details of mooring layout and deployment are given in van Haren et al. (2021). In summary, the sensors were taped at 2-m intervals to 45 vertical lines 125-m tall that were each tensioned to 1.3 kN by a single buoy on top. The T-sensors recorded data at an interval of 2 s. Three buoys held a 2-MHz single-point Nortek AquaDopp acoustic current meter recording at a rate of once per 600 s waterflow and relative acoustic intensity dI = (amp - $I_{ref}$)*0.45. Raw amplitudes 'amp', taken relative to a background value $I_{ref}$, are transferred to dI with units of dB (decibels). For somewhat better comparison with volume backscattering, dI are transferred to echo intensities 'rEI' with arbitrary units of (backscatter) power, rEI = $10^{dI/10}$. Both dI and rEI are used, although the former favours low- over high-frequency signals compared to the latter.

The lines were attached at 9.5-m horizontal intervals to a steel-cable grid that was tensioned inside a 70-m diameter steel-tube ring (Fig. 1b). Filled with air, the ring functioned as a large float that was towed to the mooring site. Filled with water, the ring was a 140-kN anchor. The ensemble 'large-ring mooring' was deployed at the <1° flat and 2458-m deep seafloor of 42° 49.50′N, 006° 11.78′E just 5 km south of the foot of the steep continental slope of the Northwestern Mediterranean Sea, in October 2020. With the help from Irish Marine Institute Remotely Operated Vehicle (ROV) 'Holland I', all 45 vertical lines with T-sensors were cut and successfully recovered in March 2024.

As with all NIOZ4 T-sensors (van Haren, 2018), their individual clocks were synchronised via induction to a single standard clock every 4 hours, so that all were sampled within 0.01 s. One line did not register synchronisation, possibly due to an electric cable failure. Three T-sensors leaked and <10 were shifted in position due to a tape malfunctioning. In total 2902 out of 2925 T-sensors functioned as expected mechanically. Due to unknown causes all T-sensors switched off unintentionally when the file-size on the memory card reached 30 MB after 7500 start-ups for writing a 4-kB data block. It implied that a maximum of 20 months of data was obtained. After post-processing, some 20 extra T-sensors are not further considered due to electronics (noise) problems. Depending on the period of investigation, between 50 and 150 T-sensors are not considered because of bias. Data from T-sensors failing post-processing criteria are interpolated between neighbouring sensors.

With respect to previous NIOZ4 version, the somewhat improved electronics resulted in about twice lower noise levels of 0.00003°C and twice longer battery life. As detailed elsewhere (van Haren, 2018), calibration yielded a relative precision of <0.001°C. Instrumental electronic drift of typically 0.001°C

mo$^{-1}$ after aging was corrected by referencing daily-averaged vertical profiles, which must be stable from turbulent overturning perspective in a stratified environment, to a smooth polynomial without instabilities. For reference and to establish a temperature-density relationship, a single shipborne Conductivity-Temperature-Depth profile was measured locally during the deployment cruise. In addition because vertical temperature (density) gradients are so small in the deep Mediterranean, reference was made to periods of typically one hour duration that were homogeneous with temperature variations smaller than instrumental noise level (van Haren, 2022). Such periods were found on days 350, 453 and 657 in the existing records. A tertiary correction involved low-pass noise filtering 'lpf' of data, with cut-off frequencies between 700 and 3000 cpd (cycles per day) and 0.05 and 0.2 cpm (cycles per meter).

## 4 Results

The 600-d data-overview time series in Fig. 2 demonstrate multiple variations with time of which a variation of 10-20 days stands out. This mesoscale variation is apparent in temperature at all levels between h = 1 and 125 m above seafloor (Fig. 2a) and in echo intensity measured at h = 126 m (Fig. 2b). It is also evident in the modulation of band-pass filtered sub-mesoscale (between [0.1 1] cpd) and near-inertial (between [0.95 1.05]f) waterflow amplitudes (Fig. 2c), in 45±5-m horizontal waterflow differences and stratification rate (Fig. 2d), and in wind-speed squared 'wind-load' (Fig. 2e).

The time series in Fig. 2d of mesoscale low-pass (<0.1 cpd) filtered effective inertial frequency (2) oscillates around planetary vorticity f by about ±1f. This seems a large value for the deep sea, but it corresponds well with values reported for near-surface eddies (e.g., Fine et al., 2018; Yang et al., 20219). It has a 600-d mean value of $<f_{eff}>$ = 0.91±0.9f, and anticyclonic relative vorticity slightly dominates over cyclonic vorticity. The $f_{eff}$ seems to follow stratification by 4-5 days. Mean N averaged over 124 m and 600 days is $<N>$ = 1.0±0.6f ≈ 1.1$<f_{eff}>$.

Typical range values that govern the deep-sea dynamics are: Temperature variations of 0.005°C in time and 124-m in the vertical that are alternated with periods when vertical differences are <0.0002°C, and total waterflow amplitudes are 0.05 m s$^{-1}$, which is also about the maximum value for 50-m horizontal flow differences. The threshold value of daily-mean 0.0002°C for 124-m vertical temperature difference delineates near-homogeneous (NH) conditions, difference < threshold, from stratified-water (SW) conditions, difference > threshold. Spectral analyses partially focus on 600-d averages, and partially on the distinction between these two conditions, represented by two periods of 17 days long, which is about the longest consecutive period under a single condition in the 600-d record.

**4.1 Statistical improvement?**

Prior to exploring various spectral analyses including vertical and horizontal coherence and cross-correlations, the mooring-array's improvement of statistics in reducing spectral uncertainty is explored. Fig. 3a displays spectra for a period of 17 days under SW-conditions, varying from nearly unsmoothed, except for application of a single modified-Kaiser taper window (Parks and Burrus, 1987), to a range of different averaging. The single T-sensor's spectrum has a large statistical-uncertainty width. This apparent variance variation is an optical illusion due to the logarithmic plotting of the x-axis; the statistical uncertainty is equally spread over all frequencies.

The equally-spread statistical uncertainty is substantially reduced, spectrally smoothed, when the single T-sensor's spectrum is averaged with that of data from all remaining independent T-sensors of the same mooring line. However, further reduction in variance variation by averaging with that of data from other vertical lines, or over longer more variable periods, is not equally spread over frequencies.

Most-improved statistics upon smoothing using multiple independent data records is found near the Nyquist frequency, where the smoothing closely matches that of a quasi-random distribution of spectral values (Jenkins and Watts, 1968). Around $\omega = f$, increased smoothing hardly reduces spectral uncertainty when data from multiple T-sensors are averaged. For the smoothed spectra, the width of variance variations is narrowest, most reduced around $\omega_{nar}$ = 350 cpd. Apparently at $\omega > \omega_{nar}$, incoherent, isotropic motions of equal-sized components [u, v, w] are found for which spectral variance complies with quasi-random distributions. Consequently at $\omega < \omega_{nar}$, more coherent, anisotropic motions are expected.

The effect of T-sensor bias on spectral content is small under SW, due to relatively large temperature variance, but reflects a slight diversion of the spectral slope starting at about $\omega_{nar}$. This is demonstrated after application of a vertical lpf$_z$ with cut-off at 0.1 cpm. Further smoothing involving all independent T-sensors from all 45 lines changes the spectral width over that of one line only by <30% for the low-frequency range ω ~< 30 cpd, while reducing by 85% ≈ (1 – 45$^{-1/2}$)/100 for ω > ω$_{nar}$ commensurate with random statistics. Only spectral-band smoothing, the averaging of variance of neighbouring frequency bands, substantially reduces variance variations by a factor of two around ω = f. However, band-smoothing does not treat the T-sensors as independent instruments, but rather their mean spectral values as independent per frequency band.

The effect of short-term bias removal via lpf$_z$ is more important, and more necessary, under NH than under SW. The effect is best visible in a coherence spectrum (Fig. 3b). The example for 10-m vertical coherence demonstrates a shift by maximum half an order of magnitude towards higher frequencies for filtered records compared to unfiltered records.

**4.2 Lag-correlation analysis**

Yearlong lag-correlation analysis has been performed between vertical temperature difference and various quantities. Normalized correlation values are shown in Fig. 4. Wind speed correlates reasonably with temperature with an advance of about 5 days (Fig. 4a). This same advance is observed for mesoscale relative vorticity in (anticyclonic-dominated) -f$_{eff}$, or a delay of about 4.5 days for cyclonic-dominated f$_{eff}$ (Fig. 4d). As a result, temperature (difference) and mesoscale f$_{eff}$ are in quadrature, while wind speed is in phase with -f$_{eff}$ and in anti-phase with f$_{eff}$. Approximately in-phase and more correlated with temperature (difference) are sub-mesoscale waterflow speeds, with a small advance of 0.7 days (Fig. 4c), while in anti-phase and less correlated are near-inertial waterflow speeds, with a delay of 1.2 days (Fig. 4b). Strictly in-phase with delay 0 is most correlated relative acoustic echo intensity (Fig. 4e), which implies that small particles are transported by dominant convective motions of active (temperature) scalars, or the acoustics reflect turbulence intensity.

The inference of the lag-correlation analysis is that acoustic echo intensity may be used as a, noisy, proxy for temperature differences. They directly reflect sub-mesoscale motions, but inversely near-inertial motions that are more associated with near-homogenous conditions under anticyclonic-dominated mesoscale relative vorticity that is directly forced by wind.

Whilst the temperature differences and zero-lag associated signals reasonably correlate with wind at the sea surface, the 5-day delay suggests a baroclinic response. Most puzzling is mesoscale relative vorticity, with approximately 5-day advance for anticyclonic-dominated motions suggesting a barotropic response to wind, and 5-day delay for cyclonic-dominated motions suggesting a delayed baroclinic response. As near-inertial motions associate with anticyclonic mesoscale motions, a trapping of the former by the latter is suggested as proposed by Kunze (1985). Sub-mesoscale and near-inertial motions are in quadrature and do not dominantly co-exist. As a result, mesoscale and near-inertial motions act like wind-driven sources in the deep-sea interior, with a transfer of energy to sub-mesoscale and possibly non-inertial internal wave motions.

**4.3 An indication for possible sources of energy**

Prior to the investigation of spectral power-laws, some information is extracted from limited cross-spectral information (Fig. 5). The exercise is limited because we only have waterflow measurements at h = 126 m above seafloor, which can be correlated with nearest T-sensor at h = 125 m, and because the waterflow measurements reach noise levels at $\omega > 3$ cpd, thereby barely resolving the IGW-band [$\omega_{min}$, $\omega_{max}$] for overall mean N = 1.35f. Nevertheless, the result is reasonably consistent between the three possible U-T pairings, of which only the least-resolved vertical one is shown, in average for the 600 days of T-sensor observations (Fig. 5).

All three cross-spectra wT show multiple sign changes at sub-mesoscales and clear negative vertical fluxes around 0.07 cpd at mesoscales and about 0.15 cpd at large sub-mesoscales. Differences between the three sets of instrumentation exist, but consistency is found around these frequencies. Assuming that the 600-d average is dominated by SW conditions with stable positive vertical temperature gradient, negative wT implies downgradient flux contribution, at mesoscales and lage-sub-mesoscales. Ambiguously-directed down- and up-gradient fluxes are found at higher-frequency sub-mesoscales, 0.3 cpd < $\omega$ < $\omega_{min}$, and IGW frequencies, $\omega_{min}$ < $\omega$ < $\omega_{max}$, albeit with net upgradient fluxes for about 0.3 cpd < $\omega$ < f.

Downgradient fluxes transport heat from larger to smaller scales such as in the turbulence inertial subrange, and here suggest large sub-mesoscale motions as a dominant source of energy. Counter-(up-)gradient fluxes are interpreted as restratification, and here suggest mesoscale, large sub-mesoscale and IGW (near-inertial motions) partially acting as source of energy. Countergradient internal-wave band fluxes were previously observed above steeply sloping topography for 2-day (van Haren et al., 1994) and 20-day (Gemmrich and van Haren, 2002) average data.

**4.4 Power spectra under SW- and NH-conditions**

At the large-ring mooring site, temperature variance and coherence spectra are rather featureless (Fig. 3a). No distinct spectral peaks, such as at f or at semidiurnal frequency $2\Omega$ stand out, and no spectral gaps are found. This implies that all signals distribute their energy over rather broad frequency ranges, instead of over narrow ranges like deterministic signals do, such as tides. An exception may be near-inertial kinetic energy KE(f). It also implies that the broad-range signals may be modelled by transitions over a certain frequency range after establishing their power-laws. Transitions and directionality of energy transfer will depend on the value of p.

Before continuing with 600-d average investigations, a spectral comparison is made of 17-day records under distinct SW- and NH-conditions (Fig. 6). Their mean temperature spectra differ by almost two orders of magnitude in variance, but are otherwise similarly shaped at first glance. Because temperature variance near the Nyquist frequency is still larger under SW conditions, albeit by only half an order of magnitude, the SW spectrum has not reached instrumental noise levels.

In none of the spectra a steep slope $\omega^p$, $p < -3$, is observed that represents the viscous dissipation range before noise level, as was observed for $\omega >\sim 3000$ cpd in hundred-times more stratified NE-Atlantic waters over steep topography (van Haren et al., 2016). The increase from half to two orders of magnitude difference between the deep-Mediterranean T-spectra suggests some variation in spectral slopes under differently stratified conditions (Fig. 6), while both show on average 4.5-d delayed correlation between stratification and mesoscale vorticity in $f_{eff}$ (Figs 2,4). Under both conditions, the IGW bands show a relatively flat spectral slope, with p being about zero judging from the not band-smoothed spectra, and relative peaks near their IGW-bounds.

Under NH conditions, stratification is very weak and occasionally unstable, with mean $N \approx 0.5f$ from reordered data, so that the IGW bounds shift to lower frequencies and cover a wider frequency range compared with SW conditions. At super-buoyancy frequencies, NH's T-spectrum tends to follow BO-scaling $p = -7/5$, especially clear between about $50 < \omega < 400$ cpd, before roll-off to noise. No KO-scaling is found in this spectrum, which suggests that convection turbulence dominates over shear turbulence. Between $\omega_{max} < \omega < 50$ cpd the spectral upper-bound variance tends to follow $p = -2$, the spectral slope of IW-scaling or finestructure. For most of this frequency range, internal waves cannot be freely propagating, unless very thin, < 2-m, layer stratification is dominant, which is unlikely. Despite the rather flat spectral slope in the IGW-band, the overall slope of the sub-mesoscale range tends to follow $p = -7/5$, but uncertainties are rather large.

Under SW-conditions, the IGW band (for mean $N = 2.2f = 2.0f_h$ calculated from reordered data) shifts to slightly higher frequencies compared to NH. The mean stratification equals that of minimum stratification under linear stability subject to large-scale destabilizing shear so that local neutral stability exists in the direction of Earth's rotational vector (van Haren, 2008). Nevertheless, spectral slopes are quite similar to that of NH, which may reflect that over time, on average, $f_{eff} \approx N$ so that local neutral stability exists in slantwise direction with nonlinear stability subject to small-scale shear, under both conditions. At super-buoyancy frequencies $30 < \omega \lesssim 800$ cpd the spectral upper-bound variance falls off with about $p = -2$. For $f < \omega < 30$ cpd it slopes with $p = -5/3$ of KO-scaling, with considerable low-variance dips.

The associated waterflow measurements demonstrate KE spectra with relatively large noise levels at $\omega > 3$ cpd, or just super-IGW, super-buoyancy frequencies. Nevertheless, these spectra demonstrate a stark contrast between the two conditions at $\omega < 3$ cpd. Under NH, a near-inertial peak stands out of the noise. At sub-IGW, sub-inertial frequencies $0.1 < \omega \lesssim 0.7$ cpd, KE rises to a variance level eventually above the inertial peak. In that sub-mesoscale range, the KE-spectrum slopes with about $p = -11/5$. This would imply BO-scaling as found in the T-spectrum, but uncertainties are large. Under SW-conditions, the same sub-inertial rise is found in KE, except that it starts at higher variance and over a smaller frequency range of about $0.1 < \omega \sim< 0.4$ cpd. In this frequency range ambiguously directed vertical flux was found (Fig. 5), and elevations and depressions in KE and T alternate, consistently under both conditions (Fig. 6). No distinct near-inertial peak is observed under SW, but a broad rise exists between about 0.4 cpd $< \omega < f$, which is also observed in T-variance. It suggests a widening of IGW, possibly in a slanted frame of reference where N $\approx 0.3f \approx 0.4$ cpd. Only at f, the KE-value under SW is lower than the peak under NH-conditions. At all other frequencies, including super-buoyancy frequencies towards noise, KE under SW exceeds that under NH. This supports the suggestion of a direct link between the rise at (sub-)mesoscale across IGW into turbulence.

### 4.5 Coherence under SW- and NH-conditions

The large-ring mooring offers the possibility to investigate temperature coherence over a range of vertical (dz) and horizontal (dh) length scales (Fig. 7). We distinguish SW- from NH-conditions, and employ lpf affecting NH as for Fig. 2b, except when vertical 2-m scale coherence is computed which scale is shorter than the filter cut-off. Different smoothing is applied by including different amounts of T-sensors (lines), as indicated in the caption of Fig. 7.

The general observation is high coherence coh > 0.8, at sub-IGW, sub-inertial frequencies, and low coh < 0.2, the 95%-statistical significance or noise level, at $\omega > 3000$ cpd. At frequencies in between, a clear distinction is observed between the two conditions. Under NH, coherence uniformly transits between above high and low coherence values. Under SW except for dz = 2 m, coherence also steadies to a level of coh $\approx 0.25$ at super-buoyancy frequencies between about $20 < \omega < 1000$ cpd, depending on length scale, before rolling off to noise level. This is already obvious in moderately smoothed spectra.

At the coh = 0.25 level, coherence is about the same for dz and dh at identical length scales, which suggests isotropic motions at these frequencies. The coh = 0.25 small but statistically significant non-noise coherence changes hardly with length scale. It is in the frequency range with thinnest variance-variation in T-spectra after smoothing (Figs 3, 6). No such non-noise coh-level was observed in the more stratified and more turbulent observations above NE-Atlantic sloping topography (van Haren et al., 2016), but a similar observation was made in well-stratified open-ocean data (Gostiaux and van Haren, 2012). Ruling out freely propagating internal waves at super-buoyancy frequencies, interpretations, also guided by vertical phase differences, are given in terms of finestructure advection and of standing-waves (breaking) in Appendix A.

Further inspection provides some specific information on the difference between length scales and between vertical and horizontal coherence. At the smallest vertical scale of dz = 2 m, coherence is largest under SW, while that under NH is reduced by bias (Fig. 7a). Although a plateau is not reached for $SW_{dz2}$, the sliding down from it to noise level occurs between $1000 < \omega < 5000$ cpd, as may be inferred from comparison with coherence at 10-m vertical scale $SW_{dz10}$ in which a coh = 0.25 level is found.

Approximately at $\omega = 1000$ cpd, SW's low coherence level coh = 0.25 starts rolling off to noise, irrespective of the scale used (Fig. 7a-d). Its starting frequency $\omega_{sf}$ does vary with length scale, between about 30 and 150 cpd, for length scales of 60 and 10 m, respectively (Fig. 7b, c). It hardly differs between vertical and horizontal coherence, and thereby marks the transition to isotropic motions as coherence becomes direction independent. At $\omega < \omega_{sf}$, $SW_{dz}$ and $SW_{dh}$ diverge for fixed length scale, with larger coherence at the horizontal scales. This has also been observed for limited length scales above a seamount slope in the NE-Atlantic which is dominated by larger stratified conditions and shear turbulence (van Haren at al., 2016). There, the transition to isotropic motions occurred at maximum buoyancy frequency calculated using an estimated length-scale of 0.25 m. If employed here, it would imply a transition at $8^{1/2}N_{max} \approx 30$ cpd. Also here, the larger the length scale, the larger the difference in coherence, with barely high coherence coh > 0.8 for dz = 60 m at all frequencies (Fig. 7c).

Under NH, no non-noise low-coherence levels are observed, and only limited difference in coherence between vertical and horizontal scales. Motions become isotropic at $\omega_{sf}$, which is length-scale dependent like under SW. While the transition from high to low coherence is about the same for $SW_{dz}$ as $NH_{dz}$ and $NH_{dh}$ for a length scale of 10 m (Fig. 7b), $NH_{dz60}$ is significantly more coherent than $SW_{dz60}$ at $\omega < 10$ cpd. This probably reflects relatively large scales dominating convection turbulence such as by geothermal heating from below and by (slanted) eddies from above under NH-conditions. Stratification commonly reduces vertical length scales, with reduced coherence, also in the deep sea.

That SW conditions may also differ in the vertical with more stratification away from the seafloor, is reflected for 30-m horizontal length scales in upper and lower layers in Fig. 7d. In the lower layer, coherence is almost indistinguishable from that under NH conditions. It lacks the coh = 0.25 level typical for SW conditions, and indeed observed in the upper layer. It suggests that the low-coherence level is related with non-negligible stratification and stratified turbulence under SW conditions, and not reflecting convection turbulence. However, the presented example is not typical for SW conditions and depends on the height of stratification. Upper- and lower-layer coherence spectra can be the same (five-day record starting day 605), and sometimes reversed when stratification is pushed to very near the seafloor under strong convection turbulence above (e.g., 0.8-day record starting day 606.8).

**4.6 600-d average spectra**

Considering full 20-month average spectra, the distinction between NH and SW is no longer made. Instead in the vertical, coherence- and shear-turbulence provide slightly different slopes with improved statistics, not only at super-buoyancy frequencies (Fig. 8). Splitting the vertical in three layers and applying band-smoothing for super-IGW frequencies, the three temperature spectra have significantly different slopes between about $30 \approx 2N_{max} < \omega < 400$ cpd. $N_{max}$ is defined as the maximum of small-2-m-scale buoyancy frequency. This definition is somewhat arbitrary as it depends on the scale-length between sensors.

A considerable part of above high-frequency range corresponds with the transition from high to low coherence, or the coh = 0.25 level in Fig. 7. In this range, the upper-layer slope follows KO-scaling with exponent $p = -5/3$ most closely over a limited frequency range only, the middle layer BO-scaling with exponent $p = -7/5$, and the lower layer Im-scaling with exponent $p = -1$ (Fig. 8). As a result, only the most temperature-variance containing upper layer demonstrates a small inertial subrange of dominant shear turbulence that treats temperature as a passive scalar. The remainder of the data, like under NH conditions over the entire vertical range of observations, does not show an inertial subrange but a fluent transition from buoyancy (or intermittency) subrange to noise. The viscous dissipation range is not resolved in any of the records.

For the range 1-2 < ω < 8 cpd, or roughly N < ω < $0.5N_{max}$, the three vertical levels demonstrate about equal slopes with an exponent close to p = -2. Although the upper bound of this range includes small-scale internal waves and although coherence was found to be high at almost all scales there (Fig. 7), it is anticipated that some large energetic turbulent overturns affect the spectral slope. If so, the spectral slope would correspond with finestructure 'contamination', or rather variable kinematic transport of thin and thick layers passed the T-sensors.

At sub-inertial frequencies 0.1 < ω < 0.6 cpd (extended to < f for middle and lower layers), T-spectra for all three levels correspond with BO-scaling with exponent p = -7/5 (Fig. 8). As this range is not expected to be part of the buoyancy subrange of turbulence, it may associate with sub-mesoscale eddies. With reference to Fig. 5, the frequency range includes the transition from upgradient to downgradient vertical buoyancy flux, established at h = 126 m. This associates with BO-scaling of which the energy transfer between scales is ambiguous in direction, as indicated for laboratory turbulence (Lohse and Xia, 2010), and unlike KO-scaling that is strictly downgradient. At ω < 0.1 cpd, spectra tend to slope like Im-scaling. This frequency range includes minimum buoyancy frequency $N_{min}$, which is difficult to define as it depends on sensor distance and duration.

Continued BO-scaling is also observed between 0.06 cpd < ω < $ω_{min}$ (for N = 1.35f) in KE-spectra, for which exponent p = -11/5, and in relative echo intensity spectra that closely correlate with temperature differences (Fig. 4e) and in horizontal waterflow difference spectra, for which exponent p = -7/5 and which thus act like an active tracer. In limited range between 0.06 < ω < 0.15 cpd, BO-scaling is also observed in vertical waterflow component w, as p = -7/5. The limited range is due to the relatively low signal/noise ratio, the signal exceeding noise for ω < 0.5 cpd at sub-mesoscales. The BO-scaling contrasts with w-spectra open-ocean well-stratified conditions, which are flat white noise at sub-mesoscales and only exceed noise level in the IGW bulging near N (e.g., van Haren and Gostiaux, 2009).

While the waterflow measurements are limited to only three instruments at upper level h = 126 m, the slopes are significantly different from other model slopes with exponents like p = -2 and -5/3. As indicated before, the near-inertial peak/bulge in these spectra has no correspondence in T-spectra, which is not necessarily expected, but which suggests considerable redistribution of near-inertial energy in temperature. The small bulging peak in 60-m waterflow difference indicates relatively short horizontal spatial scales existing near f. While this is expected for vertical scales in well-stratified waters, it explains 1-10% in kinetic energy, or about 10-30% of waterflow amplitude. This suggests the typical length scale of inertial excursions is up to 600 m, having an amplitude of 0.03 m s$^{-1}$, as observed at h = 126 m.

As observed previously from mid-depth data (van Haren, 2025), the near-inertial peak in KE extends above the level of a bridge that slopes with exponent p = -1. Such an elevated spectral bridge in KE-spectra is expected for turbulence in unstable (atmospheric) stratification (Lin, 1969). It is attributed to the flow absorbing energy from temperature variations when potential energy is transferred to kinetic energy.

**4.7 Refining fully averaged spectra in scaled form**

600-day and 38-line average spectra of 20-s sub-sampled T-sensor data are referenced to $\omega^{-7/5}$, the BO-scaling (Fig. 9). All vertical lines are separated in two groups of T-sensors, of which the upper (Fig. 9a) is composed of one, the upper-most T-sensor record, and the lower (Fig. 9b) is composed of 21 records. Only lines that had more than 80% of good data in the group are considered. In Fig. 9a, the average relative echo intensity dI-spectrum is added for reference.

The two enlarged T-spectra are quite similar. All smoothing via averaging data of the large number of > 700 independent T-sensor records for the lower spectrum hardly reduces the spectral width in variance variation in the range between mesoscale and isotropic turbulence motions. In the same band, also the dI-spectrum hardly reduces variance by averaging data from multiple acoustic beams. Least-reduced variance variation is found in the IGW band and up to $\omega$ = 3-4 cpd $\approx N_m$. The latter frequency is close to the mean Ozmidov frequency $U/L_O \approx 3.2$ cpd, where $L_O = (\varepsilon/N^3)^{1/2}$ denotes the largest length-scale for isotropic turbulent overturns in stratified waters (Ozmidov, 1965b), for mean U = 0.035 m s$^{-1}$ and convection turbulence dissipation in (slanted) weak stratification N = 0.3f.

The lack of smoothing is attributed to quasi-coherent anisotropic motions. Here, reference is made to the non-band-smoothed spectra. The most elevated part in variance is about one-and-a-half orders of frequency range wide between 0.1 < $\omega$ < 3 cpd. The top of its variance distribution is flat, i.e. follows

BO-scaling. It possibly represents a "buoyancy subrange" of sub-mesoscale and IGW motions. This frequency range corresponds roughly with the range between 2-m scale $N_{min}$, noting difficulty in establishing its value (Section 4.8), and $N_m$.

Towards lower and higher frequencies, the elevated part in variance drops off following about Im-scaling and IW/finestructure scaling, respectively. (In theory for the band-smoothed spectra, a deviation from BO-scaling is significantly following KO-scaling when the spectral slope is maintained over half an order of magnitude, Im-scaling over at least one-quarter order of magnitude, and IW/finestructure-scaling over at least one-tenth order of magnitude.) One order of frequency range away from the elevated part, variance is reduced by about one order of magnitude. At the high-frequency side this point is reached close to 2-m-scale $N_{max}$. Further beyond these frequencies, the spectra tend to follow BO-scaling, again. There, it represents the buoyancy subrange of turbulence.

Differences between upper- and lower-group 600-d spectra are as follows. In the small IGW-band, the elevated part in variance of the upper group inclines to KO-scaling while in the lower group continues BO-scaling, before dropping off at the steeper IW/finestructure-scaling for $\omega > 3$ cpd, which appears the furthest extension resolved above noise of near-inertial KE in Fig. 8. The non-turbulence dropping off from BO-scaling is thus found very limited following KO-scaling. Likewise, at $\omega > 30$ cpd in turbulence range the upper group briefly follows BO- before KO-scaling while the lower group abruptly switches to BO-scaling for almost one order of frequency range, before rolling off to noise at $\omega > 200$ cpd. The 30-cpd transition coincides with 60-m length-scale transition to weakly-coherent isotropic motions (Fig. 7) and roughly with mean-N $\omega_O$ (Fig. 9).

Given the correspondence between the two spectra and the limited KO-scaling in the upper layer only, it appears that BO-scaling is important throughout. Because of ambiguous effect of convection-like BO-scaling on the direction of energy cascade (Lohse and Xia, 2010), the apparition of elevated T-variance part in the sub-mesoscale band may result from generation of small sub-mesoscale motions from (upgradient) near-inertial and from (downgradient) large sub-mesoscale motions. This frequency band comprises the band $0.3 < \omega < 0.6$ cpd with some positive cross-spectral contributions between w and T, resulting in a reduction in range of mean downgradient flux (Fig. 5), noting some ambiguity due to limited resolution of vertical waterflow measurements. While KO-scaling is dominated by shear turbulence and a downgradient cascade, its partial apparition in the IGW of the upper group suggests a larger spread of energy towards higher frequency internal waves and turbulence scales there.

The upper-group rEI-spectrum reflects most of above observations, with dominant BO-scaling between mesoscales and about 10 cpd, before rolling off to noise. The small elevation in the IGW and the limited range between [3, 10] cpd of p = -2-slope, in comparison with the T-spectrum, may be due to the larger influence of noise.

**4.8 The distribution of stratification**

The width between minimum and maximum buoyancy frequencies of difficult-to-smooth T-variance spectra from independent measurements reflects the core of distribution of 600 days of data of (logarithm of) 2-m vertical buoyancy frequency $N_s$ (Fig. 10). Although the distribution is less wide than the width of the quasi-coherent non-smoothed variance width in Fig. 9, possibly due to the limiting vertical scale of 2 m, it reflects the core of the coherent thick variance part and the positions of mean large-scale N, mean of maximum small-scale $N_m$, and the rare occurrence of minimum $N_{min}$ and maximum $N_{max}$ buoyancy frequencies.

As rare occurrence of rapid-fluctuating $N_{max}$ are more or less understandable, a rare occurrence of $N_{min}$ does not match its long duration as it requires persistency over at least $2\pi/N_{min} \approx 10$ days. More realistic indicators are 2.5 percentiles of the $N_s$-distribution, or a threshold curve of $\omega_{Nyquist}/\omega$ resulting in 8 and 0.5 percentiles for $N_{min}$ and $N_{max}$, respectively (Fig. 10). This places $N_{max}$ at 5-7 cpd, and $N_{min}$ at about 0.55 cpd $\approx$ 0.4f, near a dip in KE (Fig. 6) and near a bulge in w around 0.4 cpd whereby the aspect ratio is reduced to O(0.1-1) (not shown).

The lower layer shows an almost perfect lognormal distribution of $N_s$. For the upper layer the distribution is slightly skewed and also shows a small flat level, starting at about $N_m$. This level reflects more small-scale layering and hence the possibility of enhanced finestructure contamination possibly following parametric instability (cf., Appendix A).

Although a somewhat smaller vertical length scale down to about 1 m would have corresponded better with the coherent and non-smoothable parts of the spectra, the correspondence of these parts with the stratification-rate distribution indicates that the deep sea is not a pool of stagnant water. Instead, it is dynamic and variable, with sub-mesoscale eddies, internal waves, stratified turbulence operating at partially the same scales. Whilst the kinetic energy has major input at localized near-inertial frequency, a broadband response is found in the scalar temperature (and other like dI) fields. Considering that most of the spectrum follows BO-scaling with an active role for scalars, and possibly also flow vorticity or divergence, a connection is suggested between mesoscale and near-inertial motions via sub-mesoscale eddies.

**4.9 BO-scaling of sub-mesoscale motions**

The observation at all levels and all quantities, including KE and several scalars, of BO-scaling in average spectra across [0.1, 0.7] cpd where sub-mesoscale motions reside can be supported from several principle considerations.

First, coherent sub-mesoscale motions in the vertically stratified deep-sea are not associated with isotropic turbulence, so that they unlikely follow KO-scaling. Second, isotropic motions are limited by the aspect ratio of ocean basins or of the size of homogeneous layers. Commonly, the aspect ratio of motions is about unity near N and << 1 at sub-mesoscales, see also 600-d average w and KE spectra (Fig. 8), with a possible extreme shift to lower frequencies in near-homogeneous layers. As in the deep Northwestern Mediterranean near-homogeneous layers can extend up to h = 300 m above seafloor (van Haren et al., 2026) and mean sub-mesoscale flow speed amounts $U_{sm}$ = 0.013±0.009 m s$^{-1}$ (Fig. 2c), the minimum frequency $\omega_{iso}$ for isotropic motions in the lower layer is expected to be $\omega_{iso}(h) = U_{sm}/h = 0.6$ cpd. This is close to the observed dip in the KE-spectrum for NH (Fig. 6). All motions at $\omega < \omega_{iso}$ must therefore be anisotropic, even under NH (over relatively large h), although variations in aspect ratio may occur such as around 0.35 cpd. Third, our observations indicate dominant convection, both under NH, via geothermal heating from below, as well as under SW, via slanted-eddy internal wave push from above/sideways. Fourth, limited observations indicate ambiguously directed vertical fluxes in the sub-mesoscale range.

Apparently, sub-mesoscale motions are best modeled using BO-scaling, rather than KO-scaling like in Fig. 1a. The BO energy cascade may be reversible because of its ambiguous direction. This unknown, but if f-motions are trapped in anticyclonic eddies, local small-scale shear may generate marginally stable Ri ≈ 1 and non-linear motions that may lead to irreversible turbulence.

Thus a modified energy spectrum is presented for deep-sea motions in the frequency range between mesoscales and stratified turbulence (Fig. 11). Sources are at large sub-mesoscales and near-inertial frequencies (which may include tides in the open ocean), which elevate the at least one order of frequency wide ranges in between. The spectral slopes between the two sources follow BO-scaling of active scalars representing anisotropic, coherent eddy-motions. At $\omega > 2\Omega$, spectra slope with internal wave/finestructure scaling, before either adopting BO-scaling under near-homogeneous and lower-layer stratified water conditions, or KO-scaling under upper-layer stratified water conditions. The latter observation seems at odds with the condition of isotropic motions, except that, probably dominant small-scale, shear is expected to be strong under stratified conditions. This sufficiently destabilizes the apparently stable conditions.

## 5. Discussion and outlook

Generally, yearlong KE-spectra of moored waterflow observations from shallow seas and open ocean demonstrate numerous peaks at inertial, tidal and higher harmonic frequencies. In contrast, such KE-spectra from the deep Northwestern Mediterranean Sea are rather featureless except perhaps for a peak around the local inertial frequency. At the large-ring mooring, an f-peak dominates yearlong spectra, but it associates with relatively calm, near-homogeneous conditions. Under more turbulent vertically stratified water conditions such a near-inertial peak vanishes into a broader range of sub-mesoscale motions.

The deep-Mediterranean temperature-(and acoustic reflection-)scalar spectra however, demonstrate no peaks at all, regardless of stratification conditions. Their entire featurelessness points at broadband distribution of variance and draws to analyses in terms of various slope dependence on frequency. The type of instrumentation and environmental conditions limit the spectral resolution, even for the high-resolution T-sensors on the large-ring mooring.

Despite the 0.00003°C noise level the T-sensors did not resolve the viscous dissipation rate. This may be due to the sampling rate of once per 2 s, which is too slow for mm-scale turbulence dissipation motions, or due to small temperature variations down to this order of magnitude in the deep Mediterranean.

On the large scales of the spectrum, the 20 (of intended 70) months of acquired temperature data were insufficient to capture a strong convection event of deep dense-water formation. We should have left the instrumentation in the deep sea for two or more decades, perhaps at a site further in the open Provençal Basin. Nevertheless, our deep-sea observations do show extensive evidence of convection, which is either generated by general geothermal heating from below and, more energetically, by internal wave action from above. Geothermal heating is suppressed by stratification from above, and only noticeable very close to the seafloor in rare flashes (van Haren, 2026 submitted) unless conditions are near-homogeneous. The two convection processes have a typical timescale of about 15 days and thus alternate at twice that scale, which results in a broad peak around 0.03 cpd in both KE- and T-spectra. Commonly, this monthly variation is attributed to mesoscale motions, which however, do not represent single harmonic wave motions like tides.

One of the aims of the study was to investigate the potential coupling between various scales of motions, involving mesoscale, sub-mesoscale, IGW, and turbulence motions. Potential coupling may be characterized by different model-slopes with frequency that depend on frequency-range and height above the seafloor. The featureless spectra are ideal for such model-slope investigation. With the absence of resolved roll-off to the viscous dissipation range, a second surprising result is the near-absence of an inertial subrange of shear turbulence. Only in upper-layer h≈100-m, more stratified waters at super-buoyancy frequencies KO-scaling may fit observed mean spectra.

In the ocean and deep sea not only large-100-m-scale shear occurs which may dominate shear turbulence, but also small-scale near-inertial shear. This shear may dominate KO-scaling in well stratified waters. In weakly stratified waters however, BO-scaling dominates, and thus convection-turbulence with associated ambiguous energy transfer so that convection tubes reorganize before dissipating. Convection is unstable, but does not necessarily occur under neutral (shearless) conditions, and small-scale (secondary) shear is expected to be important for energy transfer.

Most of the T-spectra follow BO-scaling of active-scalar convection turbulence, not only near the seafloor under near-homogeneous conditions, but also away from the seafloor and under yearlong mixed conditions. Finestructure contamination or small-scale standing internal-wave turbulence dominate under stratified-water conditions. However, while improved statistics taking advantage of the large number of independent T-sensors support above observations in the super-buoyancy frequency range of isotropic turbulence motions, such a conclusion is not possible for mid-range frequencies between sub-meso- and maximum small-scale buoyancy frequencies. Apparently, motions at the latter frequencies fall in between near-deterministic signals like tides, for which random statistics do not apply, and near-random signals like isotropic turbulence and instrumental noise.

Obviously, the nearly 3000 T-sensors provide excellent statistics at the high-frequency end, but also reasonably at the low-frequency mesoscale side. Least improved statistics are around IGW/large turbulence frequencies, which include stratified turbulence and sub-mesoscale signals. These signals should probably be sampled over larger spatial ranges. It is found that they follow some degree of coherence at all resolved spatial scales of up to 60 m, thereby forming a bridge extending about one order of magnitude in variance over background BO-scaling. From band-smoothing it is revealed that the sub-mesoscale signals in the sub-inertial frequency range follow BO-scaling. In this range, ambiguously-directed vertical flux contributions were observed from limited waterflow measurements, but which confirm similar observations in isotropic turbulence under laboratory conditions (Lohse and Xia, 2010). The same limited waterflow measurements confirm the BO-scaling in the sub-inertial range down to mesoscales, providing a significantly different spectral slope than for temperature (and acoustic intensity), which points at active scalars.

Thus, sub-mesoscale motions may be governed by mesoscale/large sub-mesoscale and by near-inertial IGW, reflecting theoretical modeling for 2D flow of a homogeneous, nondivergent fluid with up- and down-scale cascade of kinetic energy (Fjørtoft, 1953). Thereby, the deep Mediterranean observations demonstrate a sharp contrast between mesoscale and small sub-mesoscale motions. Large (sub-)mesoscale motions act as a source, albeit with some intermittent appearance resulting in a source peak that is broader than an inertial peak. Small sub-mesoscale motions are modeled by a spectral slope of energy cascade.

The observed spectral non-KO-scaling between mesoscale and IGW, and between IGW and stratified turbulence in the weakly stratified deep Mediterranean may be verified in ocean areas. Especially the influence of (internal) tides and their potential higher harmonics following strong nonlinear interactions may deform spectra, also in the deep ocean over the rugged continental slope (e.g., van Haren et al., 2002; van Haren and Maas, 2022). In those deep-ocean KE spectra, the base of the inertial-tidal higher harmonic peaks closely sloped with exponent p = -1 over an IGW frequency range commensurate with N = 5-8f, as observed in Mediterranean data (van Haren, 2025).

For future spectral improvement, mooring duration may be extended to span >10 years underwater, to include all mesoscale motions up to basin scales. For statistical significance reduction it is suggested to continue using 3D mooring arrays of closely spaced multiple instrumentation, perhaps over a somewhat wider range of up to 1 km to resolve all IGW and sub-mesoscales. More current meters in the array would be helpful to improve relative vorticity measurements and include shear observations. For this, a considerable cost-reduction of higher-resolution deep-sea instrumentation is welcomed.

The suggested complex interactions between mesoscale and turbulence signals including small-scale standing waves, via sub-mesoscale and IGW --notably near-inertial-- motions lead to slantwise convection at many scales with up- and down-gradient energy transfers at the mooring-array site in the deep Western Mediterranean. The associated turbulence is largely sufficient for deep-sea life and locally for deep-sea circulation (e.g., Ferron et al., 2017), during two winters when deep dense-water formation was absent. Future observational studies are welcomed that further extend scale resolution, and for which instrumentation should be employed not only in the deep Mediterranean but also in the ocean where tides are stronger, over a variety of seafloor topography. More is to be learned.

## 6. Conclusions

Yearlong scalar and kinetic energy spectra from the deep Northwestern Mediterranean have been calculated from data provided by instrumentation on a nearly-half-cubic-hectometer 3D mooring array. They show that,

- The deep-sea energy spectrum may be adjusted between mesoscales and stratified turbulence, with two sources, at large sub-mesoscale and inertial frequencies. Scalar spectra are completely featureless, with a broad increase in variance across mesoscale and inertio-gravity wave IGW-frequencies.

- At frequencies away from the two main sources, BO-scaling of dominant anisotropic convective motions is observed across sub-mesoscales with alternating up- and down-gradient fluxes, under all conditions. BO-scaling characterizes sub-mesoscales. KO-scaling is not possible in this frequency range because motions do not become isotropic.

- Depending on convection- and stratification-type, BO-scaling is also partially observed at super-buoyancy turbulence frequencies, alternating with limited KO-scaling. Direct fluxes could not be established in this frequency range due to poor resolution of waterflow measurements. However, convection and shear are known to generate both turbulence and internal waves.

- Internal wave/finestructure scaling is observed at IGW frequencies up to maximum small-scale buoyancy frequency.

- Mesoscale/large sub-mesoscale and small sub-mesoscale motions are distinctly different, being energy source and energy cascade, respectively. They appear in quadrature with each other.

- Anticyclonic mesoscale motions are found in phase with winds and with near-inertial motions.

- Apparent stratified water conditions and increased sub-mesoscale activity occur around the transition from anticyclonic to cyclonic mesoscale relative vorticity.

- Stratified water conditions have $N = 2.2f$, which $\approx f_{eff}$ when averaged over the full period (of dominant cyclonic relative vorticity). The former suggests linear marginal stability for large-scale shear. The latter approximate equation suggests nonlinear marginal stability for small (likely near-inertial) shear, in which case near-homogeneous $N \approx < 0.3$ cpd probably occurs in the direction of slanted convection.

- Relative acoustic echo intensity is found to be a close proxy of temperature variations, despite more noisy instrumentation. The good correspondence at zero lag suggests either >1-mm size particles being transported by water motions that also transport different (temperature) water masses, or 2-MHz acoustics is basically reflecting off temperature difference interfaces. This may be verified with multiple frequency acoustic instrumentation.

• Both scalar properties show improved smoothing by averaging spectra from multiple instruments, albeit at super-buoyancy frequencies of weakly coherent isotropic motions only. At the scales of the array of maximum 60 m horizontally and 124 m vertically, anisotropic coherent motions reduce such smoothing, especially in IGW and sub-mesoscale bands for which band-smoothing has to be applied.

**Conflict of Interest**

The author declares no conflict of interest relevant to this study.

**Data availability**

Only raw data are stored from the T-sensor mooring-array. Analyses proceed via extensive post-processing, including manual checks, which are adapted to the specific analysis task. Because of the complex processing the raw data are not made publicly accessible. The movie to Fig. A2 can be found in van Haren (2026), "Movie to: Coupling between sub-mesoscale eddies, internal waves, and turbulence in the deep Mediterranean: A spectral investigation.", Mendeley Data, V1, https://doi.org/10.17632/dxmdv75kw8.1. Current meter data are available from van Haren (2025): "Large-ring mooring current meter and CTD data", Mendeley Data, V1, https://doi.org/10.17632/f8kfwcvtdn.1. Atmospheric data are retrieved from https://content.meteoblue.com/en/business-solutions/weather-apis/dataset-api.

**Acknowledgements** Captains and crews of R/V Pelagia are thanked for the very pleasant cooperation. I also thank the team of ROV Holland I for the well-performed underwater mission to recover the instrumentation of the large ring. NIOZ colleagues notably from NMF department are thanked for their indispensable contributions during the long preparatory and construction phases to make the unique sea-operation successful. I highly appreciated working with colleagues within the KM3NeT collaboration. I thank A. Margiotta for discussions on publication matters, oceanography, and many other topics. The author acknowledges the financial support of Nederlandse organisatie voor Wetenschappelijk Onderzoek (NWO), the Netherlands.

**Appendix A Super-buoyancy finestructure contamination and parametric instabilities**

Non-negligible, small but significant coherence at super-buoyancy frequencies like the coh = 0.25 levels under weakly stratified deep-Mediterranean SW conditions of Fig. 7 has also been observed, in slightly different form, in the well-stratified open ocean far away from boundaries (van Haren and Gostiaux, 2009). There, high-frequency super-buoyancy ''waves'' were attributed to kinematics as explained using a model of advection by the vertical N-wave motion of thin, <2.5 m mainly, and thick, average 60 m, layers including inversions passed moored T-sensors (Gostiaux and van Haren, 2012). The kinematics model did not include dynamical evolution, e.g. by turbulent overturning, of the inversions or layering.

The small super-buoyancy coherence may also be due to dynamical coupling to buoyancy frequency motions via wave-wave interactions leading to high-frequency parametric instabilities of local mode-2 standing waves, as previously observed in the laboratory (Davis and Acrivos, 1967). Such non-propagating instabilities may then be advected passed the sensors by coherent internal waves. Turbulent overturns in general and slantwise convection tubes (e.g., Straneo et al., 2002) also provide quasi mode-2 motions, also on large scales O(10-100) m.

An important criterion for distinction between the kinematics and dynamics is considering the vertical phase differences, such as in Fig. A1. While for coh > 0.3 the, vertical and horizontal, phase difference is close to zero, it spreads to ±180° phase difference for the super-buoyancy coh = 0.25 level. The transition in phase difference is gradual in the vertical while abrupt in the horizontal, before reaching random distribution towards noise levels. The phase transition, while not found at 2-m intervals (Fig. 7), depends on scale-length like the coh = 0.25 level in Fig. 7, which was found at frequencies around 400 cpd for which the temperature spectrum had a slope with exponent $p = -2$.

These observations contrast with 1D open-ocean observations (Gostiaux and van Haren, 2012), where the large 100-m scale buoyancy frequency $N = 26$ cpd $\gg f$ and smallest $\Delta z = 2.5$ m. In those data, the super-buoyancy temperature spectra had a slope with exponent $p = -8/3$, which was attributed to finite layering. A zero-coherence dip was found between low-frequency coh > 0.3 and super-buoyancy coh = 0.25, rather than a smooth transition to a non-noise level as found in the deep Mediterranean data. An abrupt transition in vertical phase difference was found at all vertical scales, with a block in transition-frequency at $N_t = 1.6N \approx N_{max10}$, the maximum buoyancy frequency at 10-m scales, for $\Delta z \geq 10$ m.

For open-ocean data a scale of 10 m seems canonical, at which a separation is suggested between internal-wave shear and turbulence (Gargett et al., 1981), and below it the temperature gradient variance rolls off (Gregg, 1977).

The 10-m scale is also applied horizontally in the set-up of the large-ring mooring, but a blocking limit is not observed in the present data in weakly stratified waters of the deep Mediterranean. A 10-m scale is also roughly determined from some time-depth images as local mode-2 motions between isotherms (Fig. A2). In the horizontal, such a scale is inferred from differences in rapid motions between neighbouring lines in the movie associated with this half-day period under SW-conditions. Probably processes other than shear induce turbulence, like convection and parametric instability.

In laboratory models, parametric instability grows under large-scale shear at wavelengths/frequencies about 5 times larger than the buoyancy frequency, which would imply between about 11 and 60 cpd under SW conditions in the deep Mediterranean data, for $5N$ and $5N_{max}$, respectively. In Figs 7, A1, the coh = 0.25 level is found between about 150 and 700 cpd at 10-m scales, and between about 20 and 300 cpd at 60-m scales. The matching range suggests 50-100 m is closer to a canonical scale for the deep Mediterranean.

Although the latter frequency range is characterized by a spectral slope with exponent $p = -2$ (Fig. 6), finestructure contamination having the same spectral fall-rate (Phillips, 1971) seems less dominant than in the stronger stratified NE-Atlantic. Also, turbulence generation in the deep Mediterranean is not only governed by parametric instabilities, but also by nonlinear motions generating larger overturns like around day 443.15 in the example of Fig. A2, and by slantwise convection induced via internal wave push and/or sub-mesoscale eddies. All phenomena result in dominant 180° phase differences across the relevant length scale.

It remains to be said that conclusions inferred from super-buoyancy temperature spectra should be taken with care in the presence of steps in the temperature profile.

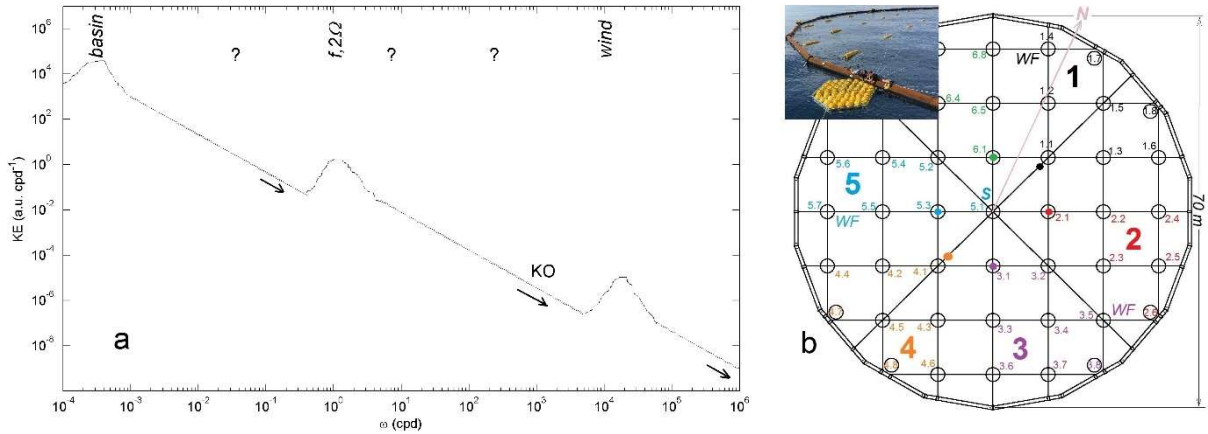

**Figure 1.** Ocean spectral model and mooring array layout. (a) Impression of ocean [kinetic] energy spectra between basin-scale motions to the left and roll-off to viscous dissipation (Kolmogorov)-scale of turbulence to the right. Redrawn from Ozmidov (1965a) and transferred from wavenumber to frequency space. The sloping lines indicate KO-scaling $\propto c\omega^p$ with exponent p = -5/3 for a model of energy downscale (direction-arrows) energy cascade by Kolmogorov (1941) and Obukhov (1949). The slopes are the same for different frequency ranges having different variance levels c. The question marks are by the author. (b) Mooring-array orientation at seabed and layout, with steel-cable grid and small rings holding the vertical lines at 9.5 m intervals. Lines are numbered in six synchronization groups. Single synchronizer S is at ring 51. Waterflow 'WF' instruments are at buoys on top of lines 14, 35 and 57. The insert shows part of the large ring just prior to deployment at sea, with free-fall drag-parachute in front.

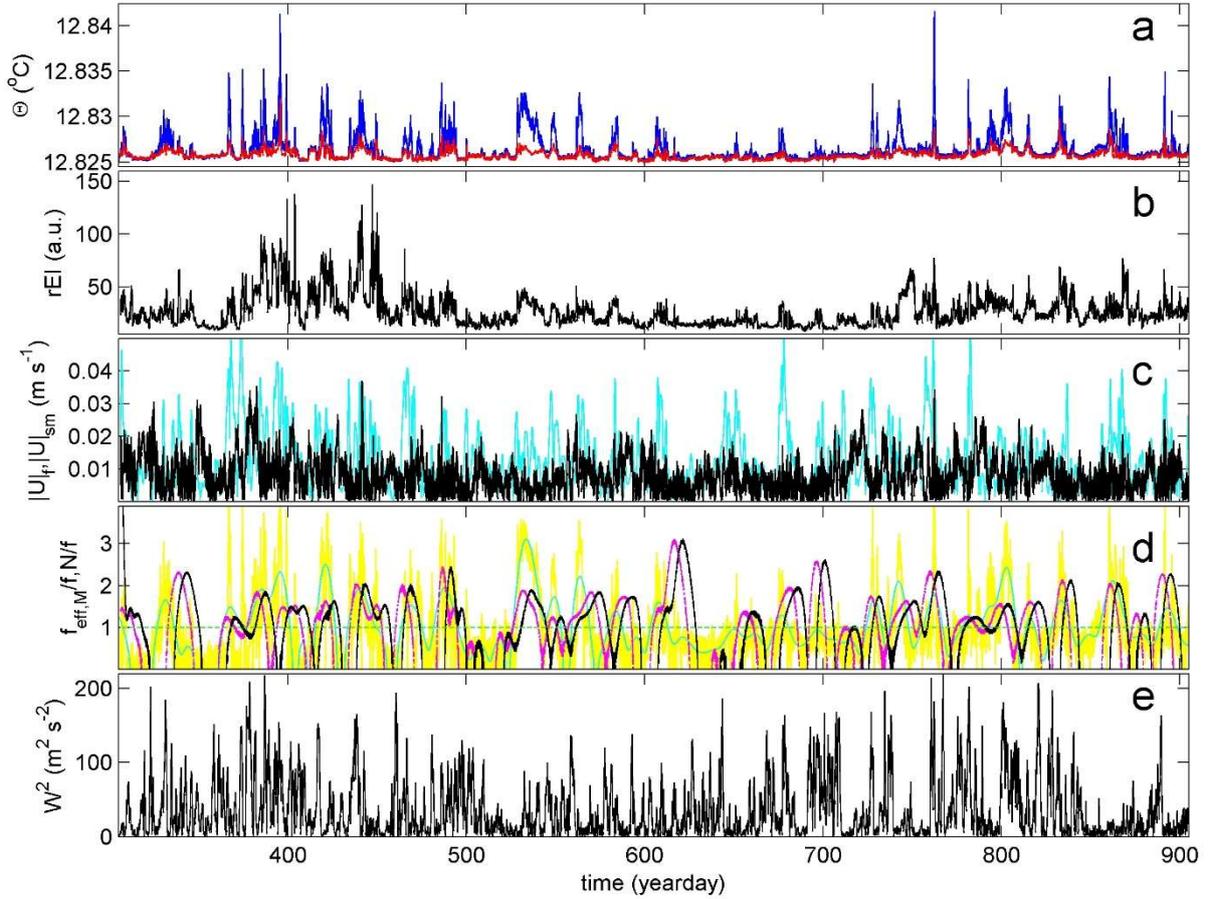

**Figure 2.** Time series of 20 months of temperature and waterflow data, (sub-)sampled at once per 600 s. Time in days starts on January 1, 2020. (a) Conservative Temperature (IOC et al., 2010) measured on a single line at h = 1.0 (red) and 125 m (blue) above seafloor. The records are detrended and referenced to shipborne CTD-data. (b) Hourly filtered echo intensity rEI measured at h = 126 m. (c) Band-pass filtered waterflow amplitudes for sub-mesoscale (cyan) and near-inertial (black) motions. (d) Mesoscale low-pass filtered effective inertial frequency (2) (black; in magenta: 4.6 days shifted) compared with 124-m scale buoyancy frequency (yellow; in green: mesoscale low-pass filtered). All are scaled by planetary inertial frequency f (horizontal dashed line). (e) Wind speed squared measured at island-station Porquerolles about 20 km north of the mooring site.

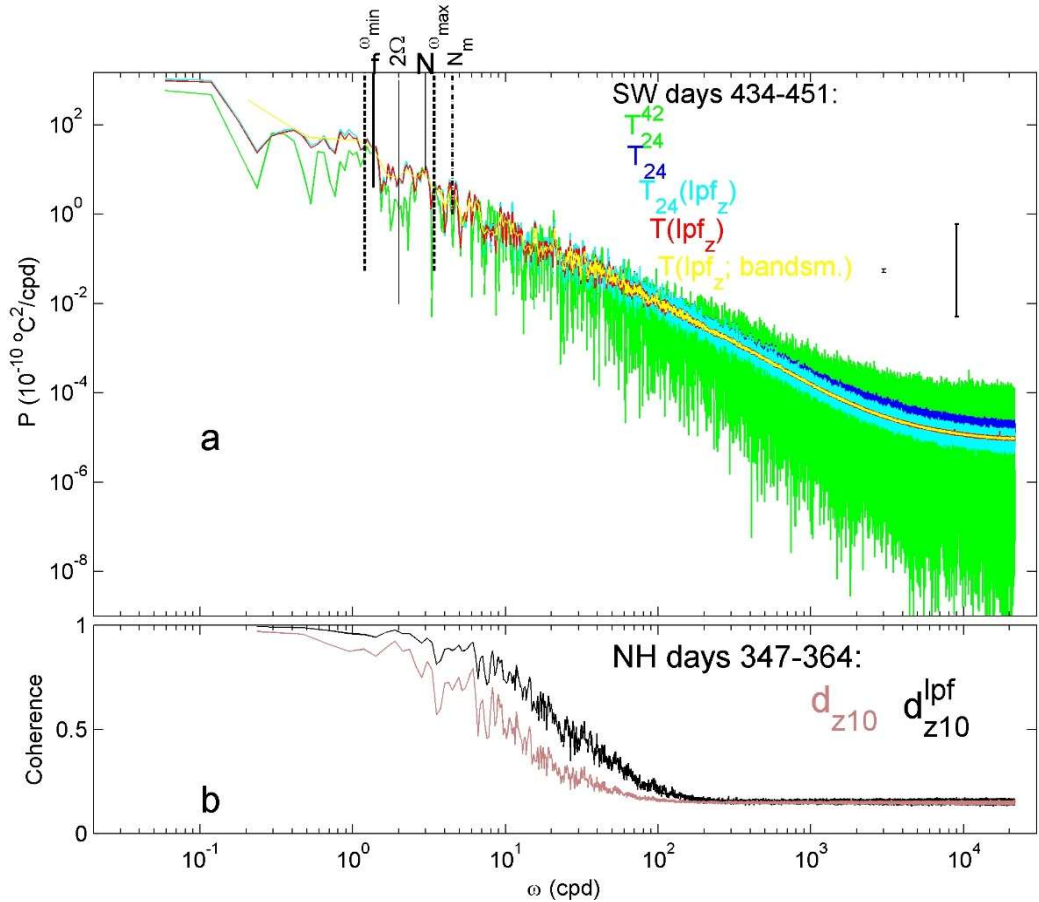

**Figure 3.** Illustration of effects of smoothing and low-pass filtering 'lpf' on 600-d mean frequency (ω) spectra from originally 2-s sampled T-sensor data. Several frequencies are named, including the inertial frequency f, the mean buoyancy frequency $N \approx 2.2f = 2.0f_h$, mean of maximum 2-m scale buoyancy frequency $N_m$, and inertio-gravity wave 'IGW' bounds [$\omega_{min}$<f, $\omega_{max}$>2Ω,N] for mean N. Ω denotes half the Earth rotational frequency. (a) Nearly raw, unsmoothed spectra of data from indicated 17-day period under stratified-water (SW) conditions. A single record (green), from 42$^{th}$ T-sensor of line 24, is compared with the vertically 63-sensor averaged spectrum of that line (blue), with the same after correction for short-term bias (cyan) using vertical lpf$_z$ with cut-off at 0.1 cpm (cycles per meter), and with the vertically and horizontally 45-line (about 2200 independent sensors) averaged spectrum (red) and its band-smoothed version (yellow). The amount of spectral smoothing is represented by small and large error bars, for the red and green spectra, respectively. (b) Coherence over vertical 10-m distances for all 1830 independent pairs of data from indicated period under near-homogeneous (NH) conditions (pink). In black, the same but after application of lpf$_z$ with cut-off at 0.05 cpm.

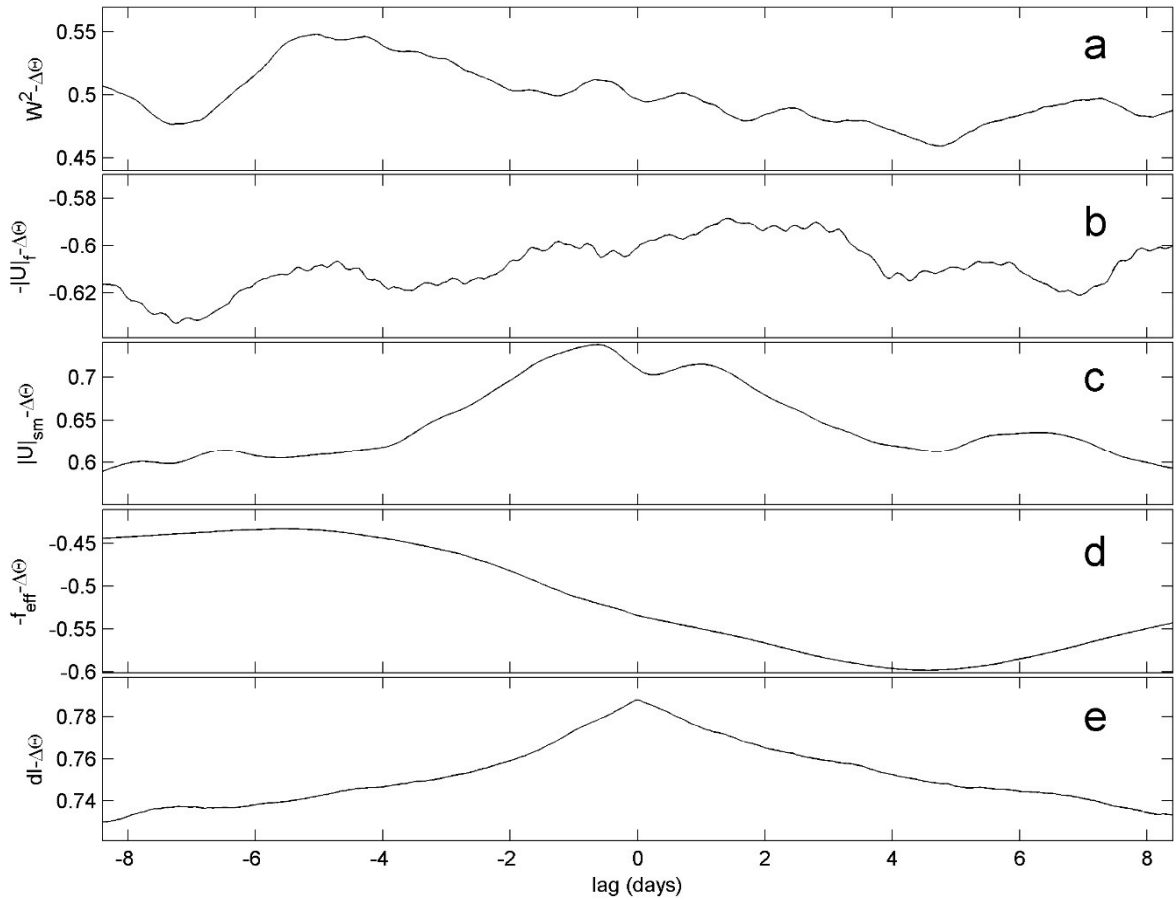

**Figure 4.** Normalized lag-correlation analysis between hourly data of h=124-m-vertical temperature difference and several observed quantities from Fig. 2. About one fortnight of lags is shown, negative lags imply that the observed quantity is ahead of temperature differences. (a) Wind speed squared. (b) Inverse (negative) inertial band-pass filtered waterflow amplitude. (c) Sub-mesoscale bandpass filtered waterflow amplitude. (d) Inverse (negative) mesoscale low-pass filtered effective inertial frequency. (e) Relative echo intensity dI(dB) averaged over 3 beams.

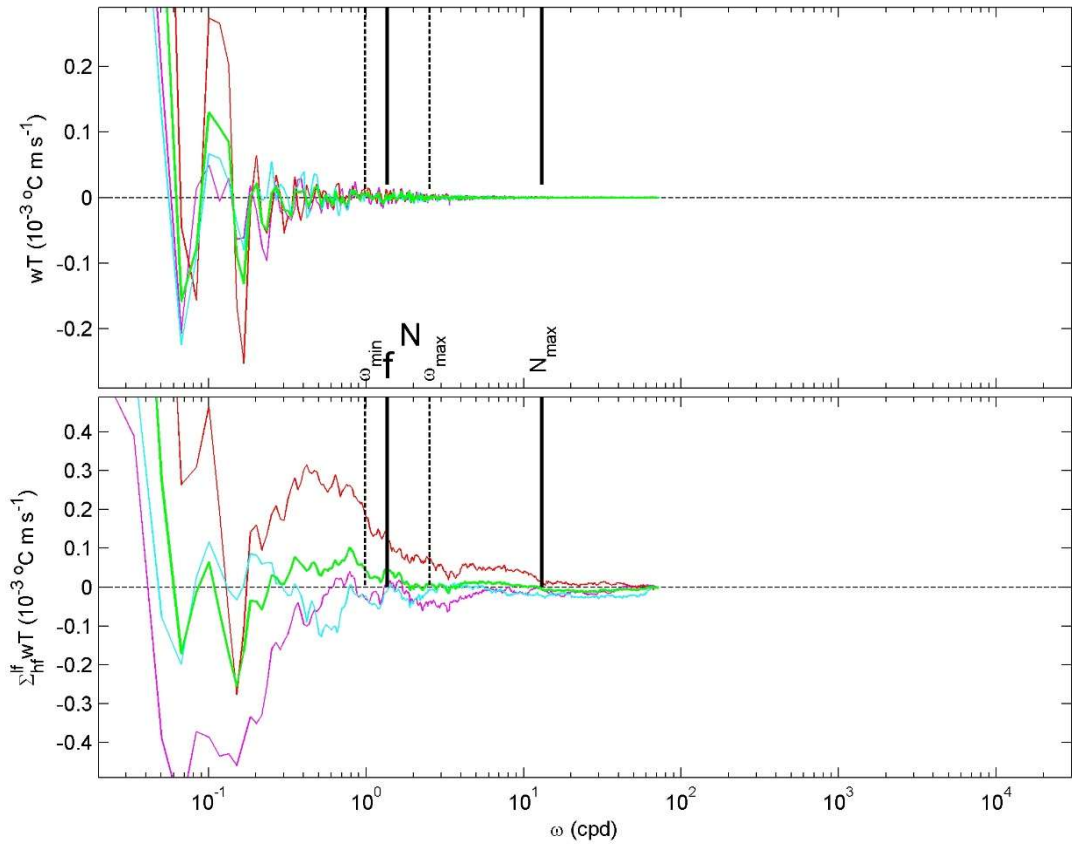

**Figure 5.** Cross-spectral averages over 600-d between uppermost T-sensor and waterflow data from all three instruments, with mean values in green. Maximum 2-m scale buoyancy frequency is indicated by $N_{max}$. The frequency range is the same as in Fig. 3, while overall mean $N = 1.35f$ is given with associated IGW-bounds. (a) Vertical cross-spectra. (b) Vertical heat flux, integrated from high to low frequencies, demonstrating the potential influence on low frequencies relative to their Reynolds decomposition cut-off.

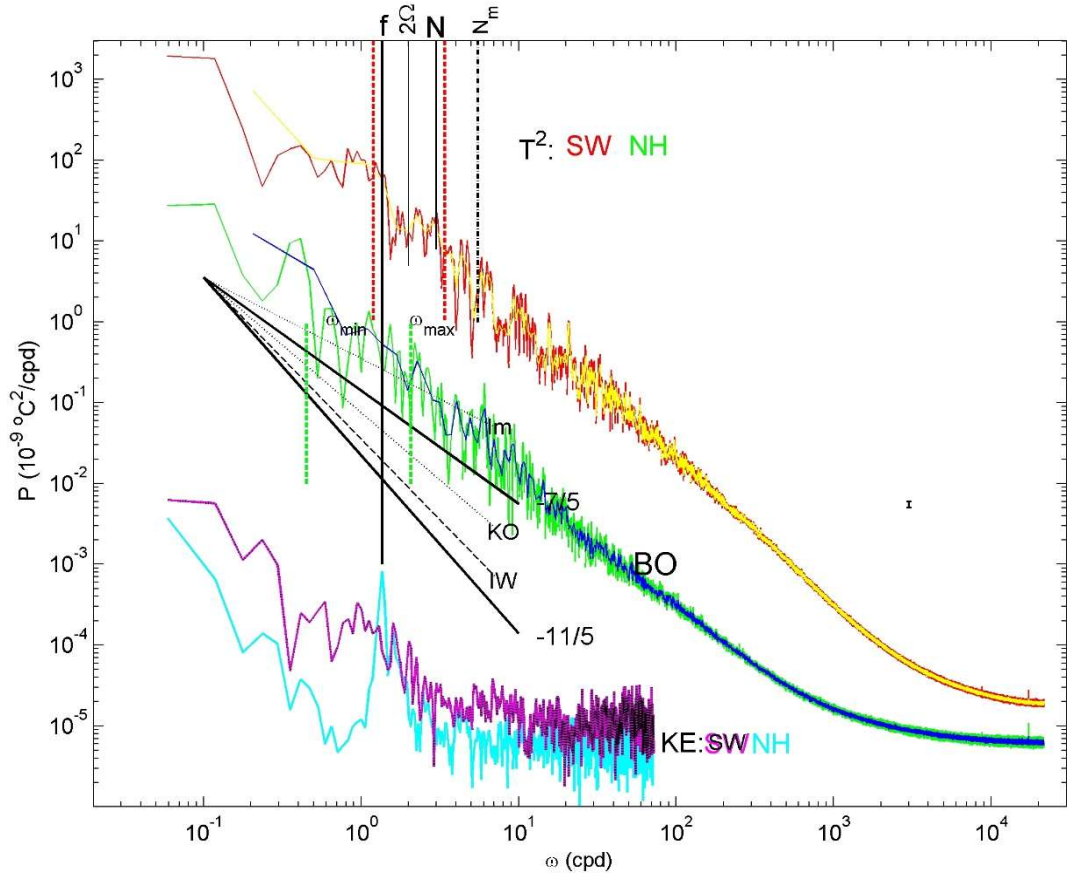

**Figure 6.** A spectral comparison between data from 17-day records of SW (red, in band-smoothed form yellow) and NH (green, band-smoothed blue) in Fig. 3, for 2-s sampled and about 2800 independent T-sensor averages in temperature. These are compared with 600-s sampled 3-sensor averages in kinetic energy KE (magenta-black and cyan, respectively). The IGW-bounds are given for mean N of 2.2f and 0.5f under SW- and NH-conditions, respectively. Several spectral slopes are indicated by their exponent value p in $\omega^p$, and by abbreviations (see text).

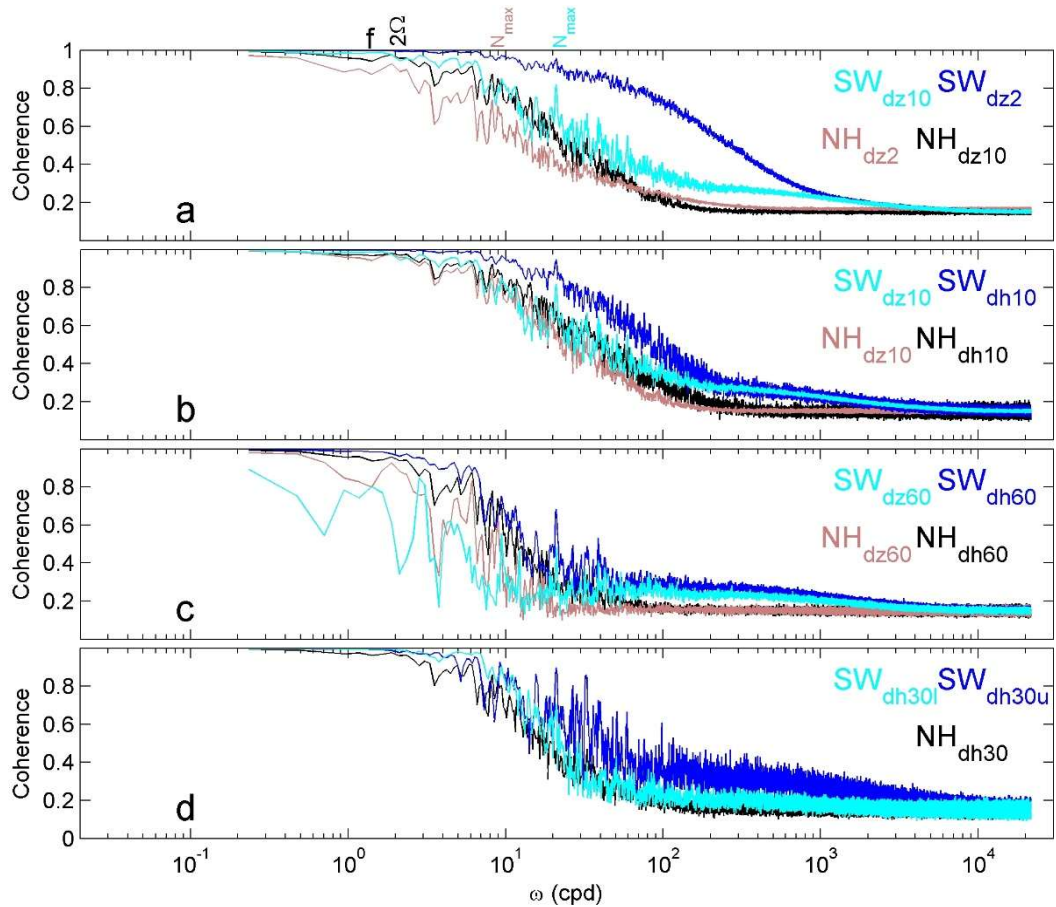

**Figure 7.** Coherence spectra using data from periods under SW- and NH-conditions in Fig. 3 for different smoothing. NH data are lpf$_z$, except for vertical dz = 2-m scale. (a) Dz 2-m and 10-m scale coherence, for about 1800 independent T-sensor pairs. Several frequencies are indicated, including maximum 2-m scale buoyancy frequency under NH and SW. (b) Ten-meter scale dz and horizontal 'dh' coherence, data from five lines around line 11. (c) As b., but for 60-m scale data of all possible 10 pairs of lines. (d) Thirty-meter scale dh for 4 line-pairs of data around the mooring-array's central line (NH) and the vertical split in three 20-sensor level groups of which upper and lower are shown (SW).

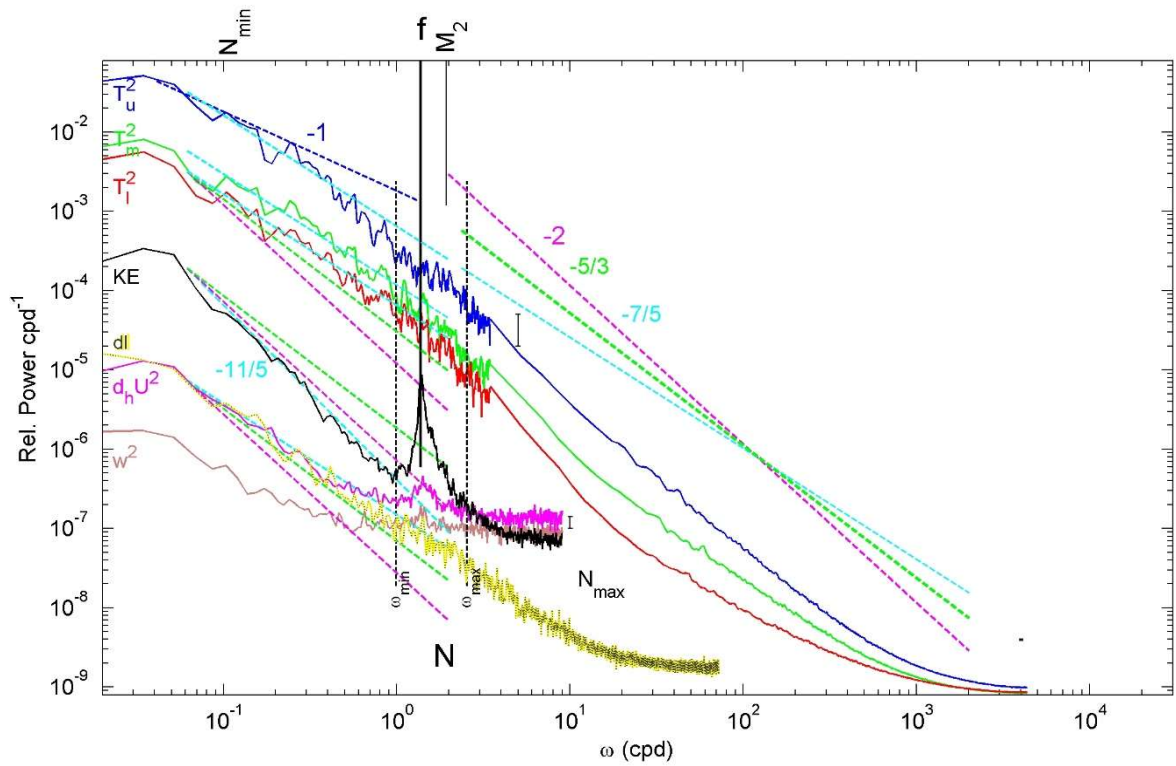

**Figure 8.** 600-d average spectra for 10-s sub-sampled data from 43 lines for four (u)pper, (m)id-height and (l)ower T-sensors. These spectra are compared with average spectra over three CM for KE, horizontal flow difference $d_hU$, vertical component w and 9-beam (3-CM) averaged relative echo intensity dI (dB). Several spectral slopes are indicated with straight dashed lines, with their exponent values. The IGW band for mean $N = 1.35f$ is between the vertical black-dashed lines. $N_{min}$ denotes the minimum 2-m scale buoyancy frequency. Each temperature spectrum is split at about 3.5 cpd (cycles per day) in a moderately-smoothed low-frequency part and a heavy-smoothed high-frequency part.

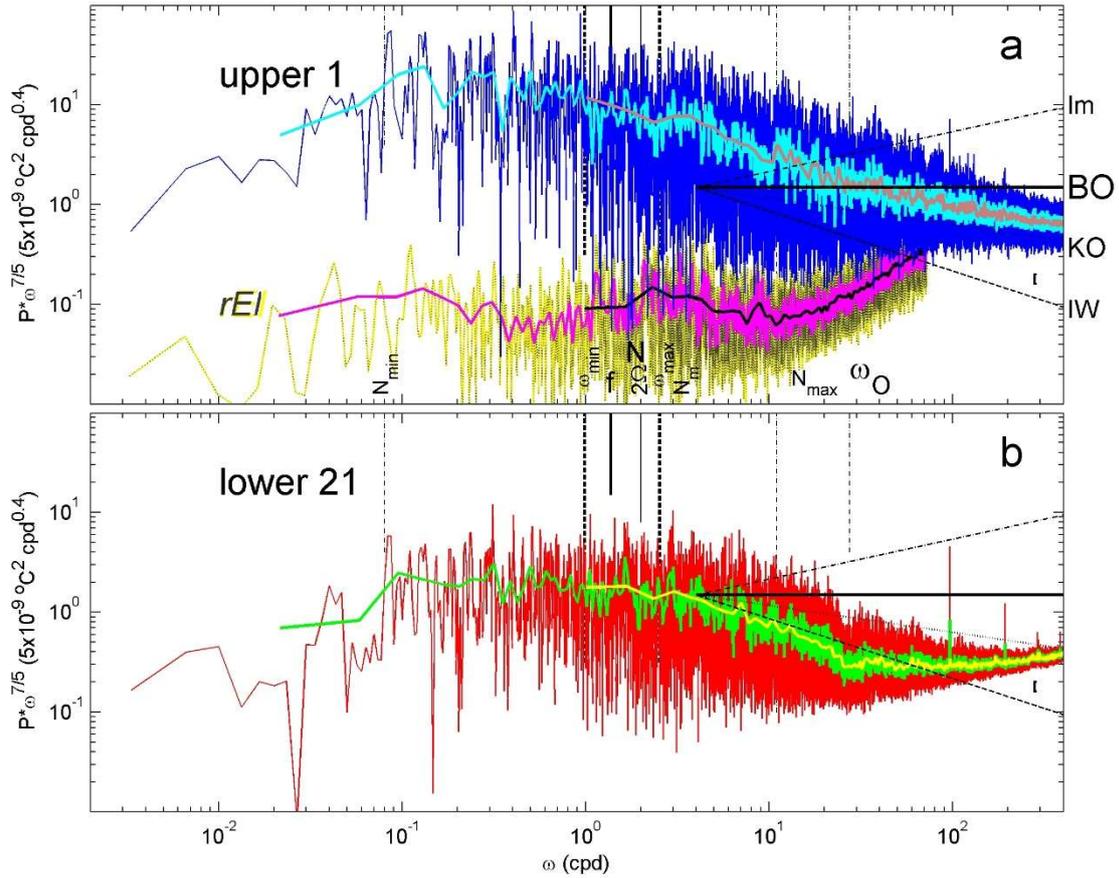

**Figure 9.** Zoom centering on IGW of 600-d and 38-line average spectra for 20-s sub-sampled data from (a) upper 1 and (b) lower 21 T-sensors. In a., the upper T-spectra are compared with 9-acoustic-beam (3-CM) averaged rEI spectra (arbitrary vertical scale). Band-smoothing is applied in the order cyan-pink for upper blue spectrum and magenta-black for yellow-grey dI-spectrum, and in b. green-yellow for lower red spectrum. The spectra are referenced to $\omega^p$, p = -7/5, BO-scaling. Sloping lines indicate other potential scalings with p-values for unscaled plots on: intermittency Im p = -1, KO p = -5/3, IW (or finestructure) p = -2 (see text). The frequency of largest isotropic overturn in stratified waters is indicated by $\omega_O = U/L_O$ for mean waterflow speed U and length scale $L_O$ (Ozmidov, 1965b). The x-axis range differs from that in previous spectra.

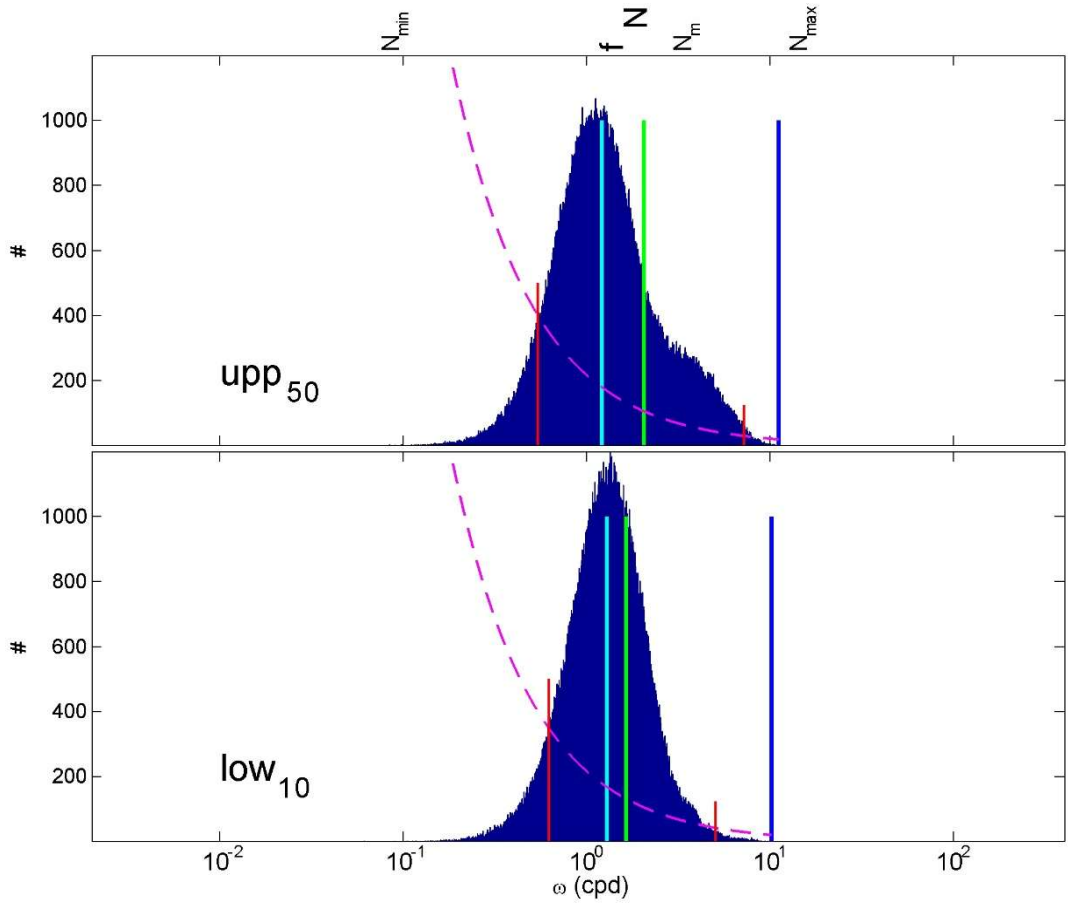

**Figure 10.** 600-d distribution of logarithm of 2-m-small-scale buoyancy frequency around T-sensors 50 (h = 99 m) and 10 (h = 19 m) from a single line. The vertical coloured lines indicate the distributions' median (cyan), mean (green) and maximum (blue) values. The dashed magenta curve indicates a threshold number of values > time/interval, i.e. > $\omega_{Nyquist}/\omega$. The x-axis range is the same as in Fig. 9.

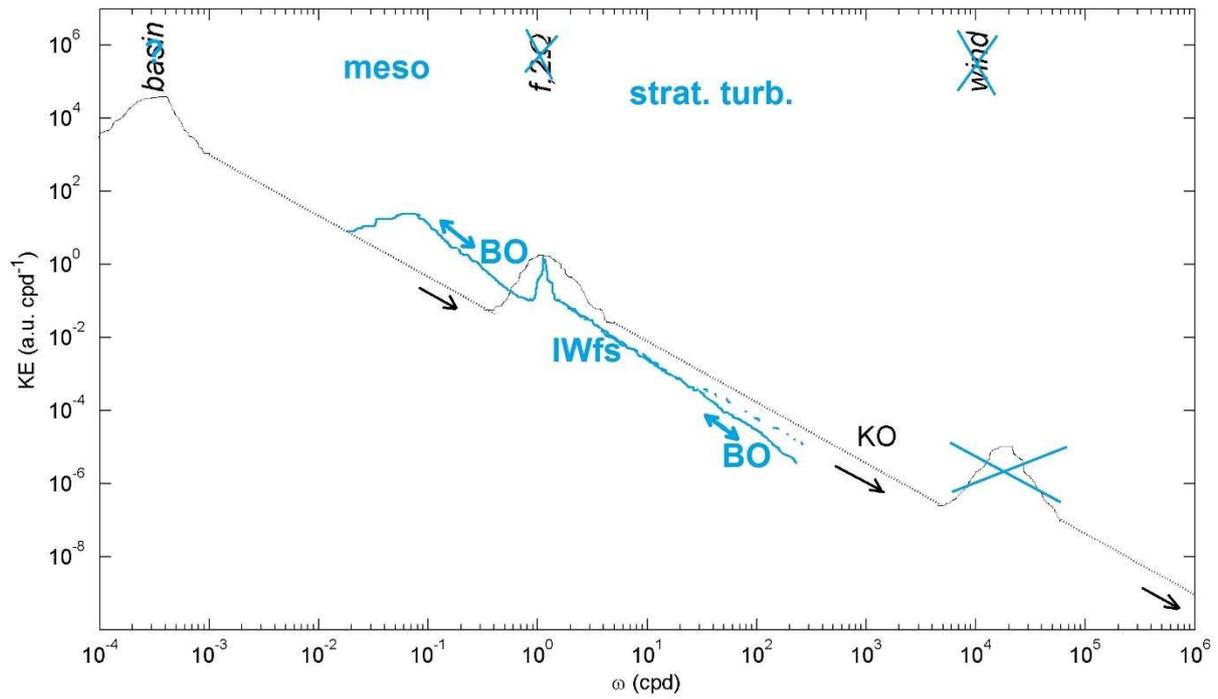

**Figure 11.** As Fig.1a, but with added information inferred from Figs 8, 9's deep Mediterranean Sea observations in heavy-solid blue lines and lettering. Double arrows indicate energy cascade direction associated with BO-scaling. (In the deep-sea wind waves are not observed.)

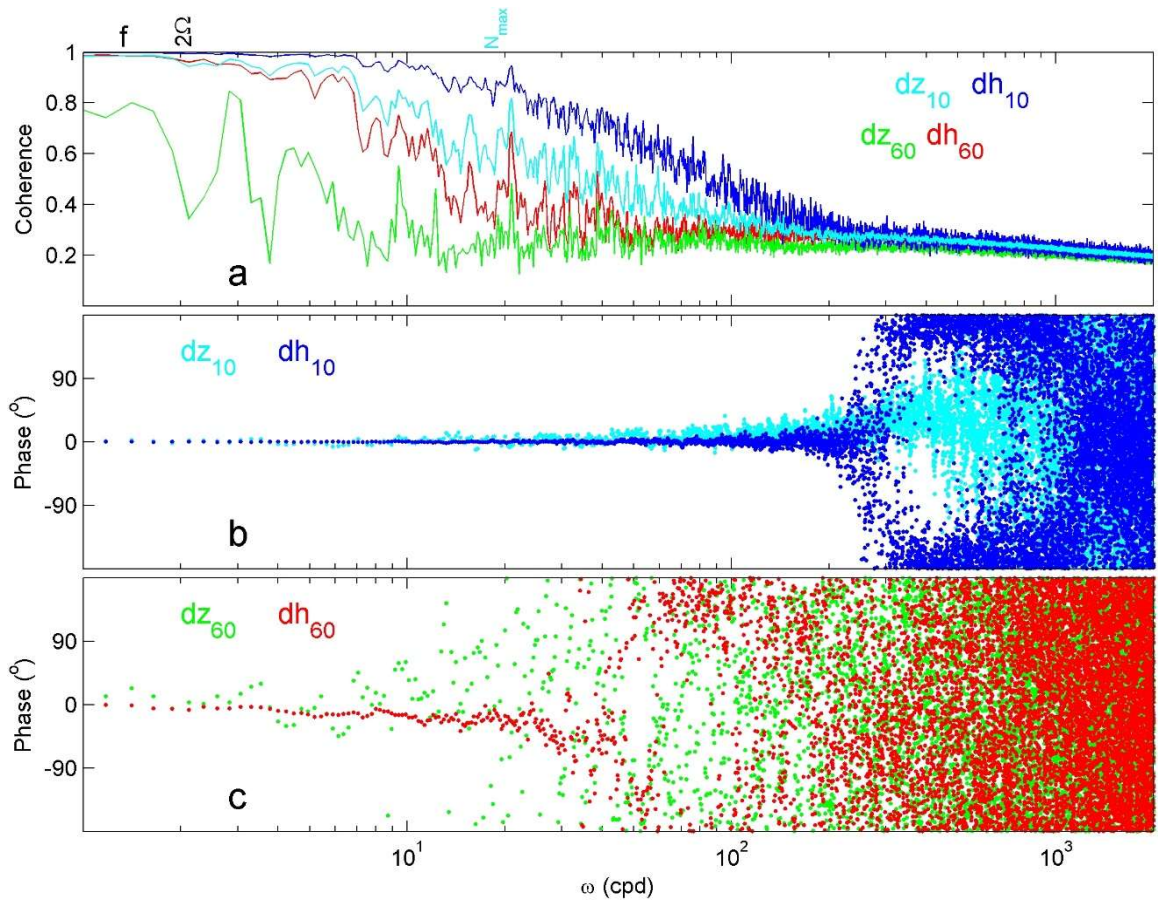

**Figure A1.** Similar to Fig. 7, but magnifying the internal wave / turbulence range and including phase differences for 10- and 60-m scales, under SW conditions only.

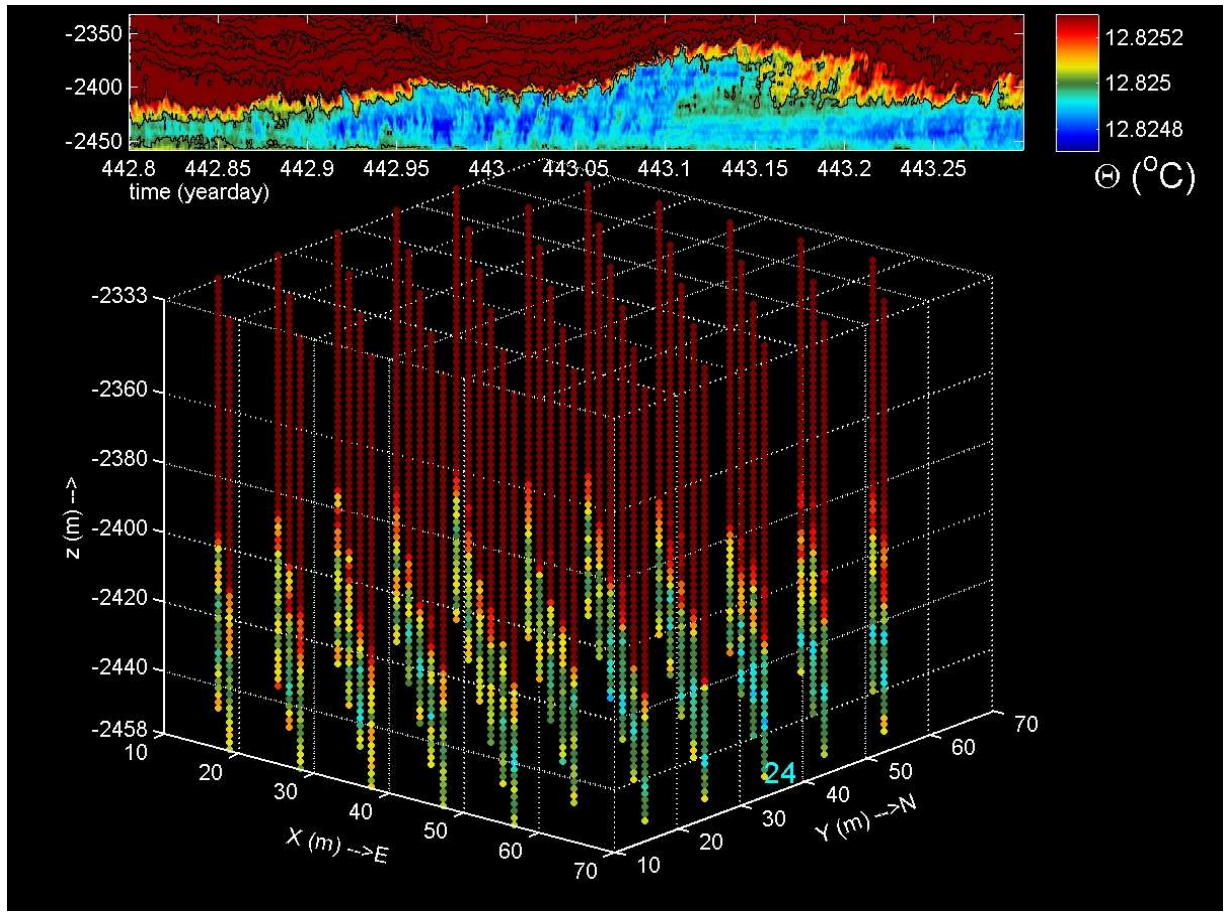

**Figure A2.** Half-day quasi-3D movie of 3000-cpd lpf temperature data from about 2800 T-sensors under SW conditions. In the cube, which is vertically depressed by a factor of two relative to horizontal scales, each sensor is represented by a small filled circle of which the colour represents Conservative Temperature in the scale above. In the movie's upper panel, a white time-line progresses in a 0.5-d/124-m time/depth image from line 24 on the east-side of the cube. Black contour lines are drawn every 0.0002°C. The 72-s movie is accelerated by a factor of 600 with respect to real-time.